\documentclass[preprint,prd,aps,floats,nofootinbib,floatfix]
{revtex4-2}

\renewcommand\i{\iota}

\newcommand{\diracslash}[1]{#1\llap{/\kern2pt}}

\newcommand{\be}{\begin{equation}}
\newcommand{\ee}{\end{equation}}
\newcommand{\bea}{\begin{eqnarray}}
\newcommand{\eea}{\end{eqnarray}}
\newcommand{\ba}[1]{\begin{array}{#1}}
\newcommand{\ea}{\end{array}}

\newcommand{\bt}{\begin{tabular}}
\newcommand{\et}{\end{tabular}}
\newcommand{\Tr}{{\rm Tr}}

\newcommand{\beas}{\begin{eqnarray*}}
\newcommand{\eeas}{\end{eqnarray*}}

\DeclareSymbolFont{rsfs}{U}{rsfs}{m}{n}
\DeclareSymbolFontAlphabet{\mathrsfs}{rsfs}

\usepackage{graphicx}

\usepackage{multirow}
\usepackage{graphicx,epstopdf}
\usepackage{amsmath}
\usepackage{graphicx}
\usepackage{subcaption}
\usepackage{comment}
\usepackage[utf8]{inputenc}

\usepackage{csquotes}
\usepackage[english]{babel}
\usepackage{color}
\usepackage{hyperref}
\usepackage{cleveref}
\hypersetup{
    colorlinks=true,
    linkcolor=blue,
    filecolor=magenta,      
    citecolor=blue, 
    urlcolor=blue,
}
 
\begin{document}

% \begin{linenumbers}
 
%\title{A study of  neutral mesons bound states with nuclei }

\title{A study of  bound states of 
$\eta, \eta^{'}, D^{0}$, $\bar{D^{0}}$, $B^{0}$, $\bar{B^{0}}$, $\bar{K^{0}}$ and   $\phi$ mesons with light and heavy nuclei within chiral SU(3) model}

\author{Arvind Kumar}
\email{kumara@nitj.ac.in}
\affiliation{Department of Physics, Dr. B R Ambedkar National Institute of Technology Jalandhar, 
 Jalandhar -- 144008, Punjab, India}
 
 \author{Amruta Mishra}
 \email{amruta@physics.iitd.ac.in}
\affiliation{Department of Physics, Indian Institute of Technology Delhi, Hauz Khas New Delhi 110016 India} 
 
\def\be{\begin{equation}}
\def\ee{\end{equation}}
\def\bearr{\begin{eqnarray}}
\def\eearr{\end{eqnarray}}
\def\zbf#1{{\bf {#1}}}
\def\bfm#1{\mbox{\boldmath $#1$}}
\def\hf{\frac{1}{2}}
\def\kp{\zbf k+\frac{\zbf q}{2}}
\def\km{-\zbf k+\frac{\zbf q}{2}}
\def\hwo{\hat\omega_1}
\def\hwt{\hat\omega_2}

\begin{abstract}
In the present work we explore the possibilities of the formation of bound states of neutral pseudoscalar mesons $\eta, \eta^{'}, D^{0}$,  $\bar{D^{0}}, B^{0},  \bar{B^{0}}$
and $\bar{K^{0}}$ and the vector meson $\phi$, with the nuclei
$^{12}$C, $^{16}$O, $^{40}$Ca and $^{208}$Pb, calculating their binding energy  and absorption decay width.
To calculate the optical potentials of these mesons
in different nuclei under study, we shall use the chiral SU(3) hadronic mean field model, in which the properties of nucleons  in the medium are modified through the  scalar isoscalar fields $\sigma$ and $\zeta$ and the scalar-isovector field $\delta$.  The  scalar-isovector field $\delta$  account for the finite isospin asymmetry of different nuclei having asymmetry in number of protons and neutrons. The binding energy and absorption decay width of mesons are calculated  for the ground state and some of possible excited states of the nuclei.  In the chiral SU(3) model, the
mesons $ \eta,  D^{0}$, $\bar{D^{0}}, B^{0}, \bar{B^{0}}$ and $\bar{K^{0}}$  are observed to have significant negative mass shift upto nuclear saturation density which lead to the possibility of the bound states at least for ground states and some of excited states in case of heavy nuclei.
For the pseudoscalar singlet $\eta^{'}$ and the vector meson $\phi$ the mass shift obtained
are found to be small and bound states are not formed.
The present calculations are compared with different studies available in the field and will be useful in understanding the outcomes from different experimental facilities focusing on this area of research.
\end{abstract}
\maketitle
 
\newpage

\section{Introduction}
\label{intro}
The theory of strong interactions, quantum chromodynamics (QCD), has two well known properties: asymptotic freedom and confinement. The understanding of the low-energy non-perturbative regime of QCD poses extra challenges due to large value of coupling constant. The properties of QCD such as breaking of chiral symmetry and its expected restoration at high temperature and baryon density play an essential role in our understanding of strong interaction physics.  The study of medium modification of hadron properties, specially pseudoscalar and vector mesons, play a pivotal role in revealing the nature of the partial restoration of the chiral symmetry at high temperature and baryon densities.
The heavy-ion collision experiments, for example, the large hadron collider (LHC) at CERN, Switzerland  and relativistic heavy-ion collider (RHIC) at BNL, USA, explore the regime of strongly interacting matter at high temperature and low baryon densities, whereas, the future experimental facilities such compressed baryon matter (CBM) of FAIR, GSI and the Nuclotron-based Ion Collider fAcility  (NICA) at Joint Institute for Nuclear
 Research (JINR), Russia intend to explore the nature of interactions and phase transitions at high baryon densities and low/moderate temperatures.  There are extensive theoretical studies using various approaches which explore the medium modification of hadron properties at finite temperature and baryon densities \cite{Hayano:2008vn,Lee:2023ofg,Yasui:2012rw,Leupold:2009kz}.

%\cite{Aguirre:2024clo}
%\bibitem{Aguirre:2024clo}
%R.~M.~Aguirre and O.~Louren\c{c}o,
%``Pseudoscalar mesons from a PNJL model at zero temperature,''
%[arXiv:2401.14119 [hep-ph]].
%0 citations counted in INSPIRE as of 30 Jul 2024

An alternative technique which  play a significant role in understanding the strongly interacting matter and role of chiral symmetry is the study of mesic nuclei, i.e., the bound states of mesons and nuclei, where strong interactions play a dominant  role. In the formation of mesic nuclei, one of the nucleon in a given nuclei is replaced with the meson. This is complementary to mesic atoms in which mesons replaces electron in the atom
and Coulomb interactions play the key role  \cite{Metag:2017yuh}. Formation of mesic nuclei for charged and neutral mesons has been studied using different approach, for example, quark meson coupling (QMC) model \cite{Tsushima:1998ru,Cobos-Martinez:2017woo}, 
Nambu–Jona-Lasinio model (NJL) \cite{Nagahiro:2006dr}, chiral unitary  approach \cite{Nagahiro:2011fi,Garcia-Recio:2010fiq} etc.
The mesons which have strong attractive interactions in the dense nuclear medium, i.e., whose mass deceases with increase in the baryon density and thus, have deep attractive potential, may form the  bound states with different nuclei.

The study of $\eta$ and $\eta^{'}$ mesic nuclei shed light not only on interactions of these mesons with nucleons in the nuclear matter, but is also important to deepen our understanding  on the behaviour of $U_A(1)$ anomaly in the dense medium.
Initial studies in the direction of $\eta$ mesic nuclei were initiated
by  Haider and Liu \cite{Haider:1986sa}. In this work, through the analysis of $\pi N \longrightarrow \eta N$  cross-sections and solving relativistic wave equation, possibilities of existence of bound states of $\eta$ mesons with nuclei having mass number $A\geq 12$
have been claimed.
In \cite{Cheng:1987yw}, Green function method is applied to study the $\eta- ^{16}$O bound state.
Subsequent to these initial studies, in-medium properties of $\eta$ and $\eta^{'}$ mesons and possibility of their bound state formation with different nuclei have been studied by different groups  \cite{Nagahiro:2012aq,Hirenzaki:2013txr,Cobos-Martinez:2023hbp,Kelkar:2013lwa,Garcia-Recio:2002xog,Sakai:2022xao}.
The formation of $\eta^{'}
(958)-$nucleus bound state is investigated  in the $(p, d)$ reaction to explore the medium modification of
$\eta^{'}
(958)$ mass \cite{Nagahiro:2012aq}.
Many experiments are devoted to search the $\eta$ mesic nuclei by different collaborations
by studying reactions from heavy  and light targets.
Experiments at BNL \cite{Chrien:1988gn}
involved the reaction of
$\pi^+$ with  targets   lithium,
 carbon, oxygen, and aluminium, whereas, Los Alamos Meson Physics Facility (LAMPF) considered $^{18}$O \cite{Johnson:1993zy},
but both with negative results.
The photo-production process, induced by reaction of $\gamma$ photon on $^{12}$C,
 lead to the formation of $\eta-$nucleon bound state, as
claimed by Lebedev Physical Institute (LPI) group \cite{Sokol:2001hx}. 
Deuteron induced reaction on carbon nuclei was studied in
  JINR to investigate the formation of $\eta$ mesic nuclei through the decay of  $N^*(1535)$ resonance \cite{Afanasiev:2011zza}.
 Along with heavy nuclei as targets, experimentalist are actively looking for the  bound state of $\eta-$ mesons with light nuclei, for example, WASA at COSY  collaboration through reaction channels $pd \rightarrow ^{3}$He 2 $\gamma$ and 
 $pd \rightarrow ^{3}$He 6 $\gamma$ are looking for bound states of $\eta$ with $^{3}$He nuclei \cite{WASA-at-COSY:2020bch,Adlarson:2019haw}. Theoretically, using few body $\eta-NNN$ system existence of  $^{3}$He $\eta$ bound state is explored \cite{Wu:2023cfh}.
 Recent reviews on the theoretical and experimental efforts on $\eta$ and $\eta^{'}$ mesic nuclei are presented in Refs. \cite{Khreptak:2023lbh,Bass:2021rch}.

Investigation of the medium modification of open charm $D$ mesons have implications in understanding the their prouction rate, the phenomenon of  charmonium suppression which may shed light on formation of QGP, and also
in exploring the possibility of formation of $D$ \cite{Garcia-Recio:2010fiq,Tolos:2013gta} and $J/\psi$ mesic nuclei \cite{Tsushima:2011kh}.
As the open charm $D$ mesons have one light $u/d$ quark or anti-quark as its content, these mesons can undergo the significant mass modifications, through the medium modification of light quark condensates.
The medium modifications of charm mesons have been studied using QCD sum rules, chiral SU(3) hadronic model, coupled channel approach, QMC model etc. \cite{Hayashigaki:2000es,Mizutani:2006vq,Tolos:2007vh,Mondal:2023iwe,Mishra:2008cd,Kumar:2020kng,Gubler:2020hft,Suzuki:2015est,Kumar:2011ff}.
In Ref. \cite{Tsushima:1998ru}, using the QMC model, bound states of $D^-, D^0$ and $\bar{D^0}$ mesons in $^{208}$Pb were studied. Using the $D$ meson self-energies, calculated using the coupled channel approach, the bound states of
$D^0$ mesons with $^{12}$C and $^{208}$Pb were studied in Ref. \cite{Garcia-Recio:2010fiq}.
The study of $D$ meson  bound
states with the nuclei is also importrant for the $\bar{\text{P}}$ANDA experiment of FAIR project \cite{Yamagata-Sekihara:2015ebw}.
In literature the work has also been done on the bound states of $\omega$ \cite{Schadmand:2010bi}, $J/\psi$ \cite{Tsushima:2011kh}, $\eta_c$ \cite{Cobos-Martinez:2020ynh} and $\eta_b$ \cite{Cobos-Martinez:2022fmt} mesons  with different nuclei.
Although the in-medium mass of bottom meson have been studied using different approaches \cite{Pathak:2014nfa}, the $B-$mesic nuclei have not been explored except one  recent study using QMC model \cite{Mondal:2024vyt}.  

Understanding the medium modification of $\phi$ mesons is also an active topic at present due to different results obtained  from various experimental studies. 
In the KEK-PS E325 experiment, though the study of invariant mass spectra of $e^{-}e^{+}$ pairs produced in 12 GeV $p+A$ reaction, through the process  $p+A \rightarrow \phi + X \rightarrow e^{-}e^{+} + X^{'}$, a negative mass shift of $3.4$\% 
and decay width $14.5$ MeV is observed at nuclear saturation density \cite{KEK-PS-E325:2005wbm}.
Other experiments, for example, Laser-Electron Photon facility at
 SPring-8 (LEPS) \cite{Ishikawa:2004id} and 
by CLAS collaboration  JLab \cite{CLAS:2009kjz,CLAS:2010pxs}, have reported significant modifications in the in-medium  decay width, without significant effect on mass shift. The J-PARC E16 collaboration aim to investigate the mass modifications of $\phi$ mesons through the mass spectra of $e^{-} e^{+}$ pairs, as in KEK, but with more high statistics and in the energy range of 50 GeV \cite{JparcE16,Aoki:2015qla}. Theoretically, the medium modifications of $\phi$ mesons have been studied using QCD sum rules, coupled channel approach, chiral hadronic model, quark meson model etc. \cite{Hatsuda:1991ez,Hatsuda:1996xt,Klingl:1997tm,Oset:2000eg,Gubler:2015yna, Cobos-Martinez:2017woo,Paryev:2022zkt,Chizzali:2022pjd,Kim:2022eku,Kumar:2020vys}.
The negative mass shift for the $\phi$ mesons observed under various theoretical approaches and also as reported by KEK-PS 325 \cite{KEK-PS-E325:2005wbm}, motivate to study the possibility of the formation of $\phi-$ mesic nuclei \cite{Cobos-Martinez:2017woo}. 
At JPARC experimental facility, plans  are to investigate the  mass shift of $\phi$ mesons though the  $\phi$ meson-nuclei bound states using reactions $\bar{p}p \rightarrow \phi \phi$
 \cite{Japrc_nuc1,Japrc_nuc2}.

In the present paper, we shall explore the   formation of bound states of pseudoscalar mesons, $\eta, \eta^{'}, D^{0}, \bar{D^0}, B^{0}, \bar{B^0}, \bar{K^0}$ and the vector meson $\phi$, whereas, the bound states of  $\omega$, $J/\psi$, $\eta_c$  and $\eta_b$ will be investigated in future work. 
For our current work, we shall use the  chiral SU(3) hadronic mean field model which is based on low energy properties of QCD \cite{Papazoglou:1998vr}.
Chiral SU(3) model has been used in literature to study the modification of properties of kaons \cite{Mishra:2006wy,Mishra:2008kg,Mishra:2008dj},  vector mesons ($\omega, \rho $ and $\phi$) \cite{Mishra:2014rha},
pseudoscalar $\eta$ meson \cite{Kumar:2020gpu,Tiwari:2022gre},
$D$ \cite{Mishra:2008cd,Kumar:2010gb,Kumar:2011ff} and $B$ mesons \cite{Dhale:2018plh}, charmonium \cite{Kumar:2010hs} and bottomonium \cite{Mishra:2014gea},
in nuclear and strange hadronic matter. The optical potentials of mesons under study, calculated using chiral effective model for different light and heavy nuclei, will be used to solve  Klein Gordon equation
in the momentum space to obtain the binding energies and decay width.
Note that in the current work we used mean field approximation in calculating the in-medium masses and optical potentials of different mesons throughout the manuscript.
 In future, we may further improve
  the results employing beyond mean field calculations for example including two loop contribution 
 \cite{Zhang:2016bem,Constantinou:2016hvf}.

Following is the outline of present paper: In Sec. \ref{sec:chiral_model}, we shall present the details of chiral SU(3) model. Details of how to obtain the optical potentials of mesons
in the environment of nuclei will be discussed in Sec.
\ref{sec:potential}. 
The results and discussion will be presented in Sec. \ref{sec:results} and finally, 
in Sec. \ref{sec:summary}, summary and conclusion   will be given.

\section{Chiral SU(3) hadronic model} \label{sec:chiral_model}

To investigate the formation of mesic nuclei in  the present work, we shall use the chiral SU(3) hadronic  mean-field model based on
nonlinear realization of chiral symmetry 
 \cite{Papazoglou:1998vr,Weinberg:1968de,Coleman:1969sm,Bardeen:1969ra}.
In the chiral SU(3) model, the interactions between the nucleons are mediated through the exchange of   non-strange scalar meson
$\sigma$, strange scalar meson $\zeta$, scalar-isovector meson $\delta$,  vector meson $\omega$ and vector-isovector meson $\rho$.
% The $\sigma$ and $\zeta$ fields are non-strange and strange scalar-isoscalar fields.
%The  non-strange scalar meson,  $\sigma$, is composed of light $u$ and $d$ quarks, while the strange scalar meson $\zeta$ contains strange $s$ quark \cite{Zakout:1999qu}.
The scalar   field $\delta$ and vector   field $\rho$ contribute  when medium has finite isospin asymmetry.
In this model, the scalar dilaton field, $\chi$,   known as glueball field is introduced  to  incorporate the broken scale invariance property of the QCD \cite{Papazoglou:1998vr}.
% Through the exchange of these scalar fields ($\sigma, \zeta, \delta, \chi$) and vector fields ($\omega$, $\rho$ and $\phi$.), the nucleons and the hyperons in the medium interact.
As discussed in the introduction, QMC model, NJL model, chiral unitary approach, linear sigma model etc. have been used to calculate the in-medium optical potentials and hence the possibility of the formation of mesic nuclei. 
Comparing chiral SU(3) model with these approaches following are  important differences. In the QMC model, quarks are treated as fundamental degrees of freedom confined inside the hadrons through a bag potential. Quarks confined inside the baryons interact through the exchange of scalar and vector fields, which in-turn modify the properties of baryons. In the QMC model in-medium masses of kaons and antikaons
decrease as a function of density of the nuclear matter. Repulsive interactions for kaons and attractive  for antikaons contributed by Weinberg Tomozawa
term are not taken into account in this model (also for other pseudoscalar mesons).
In the chiral unitary approach, coupled channel dynamics is employed to investigate the properties of antikaons and $\phi$ mesons. Coupled channel approach involve the coupling of $\bar{K}N$ to $\pi \Sigma$ and $\pi \Lambda$ channels and is important to study the dynamics of $\Lambda(1405)$ resonance also \cite{Ramos:1999ku}. Such coupled channel effects are not considered in our present chiral model calculations.  In the NJL model  calculations for the $\eta$ and $\eta^{'}$ mesons, positive mass-shift for $\eta$ and negative mass-shift for  $\eta^{'}$ were observed \cite{Nagahiro:2006dr} . In the NJL model interaction Lagrangian densities are expressed in terms of quark fields.

The Lagrangian density for chiral SU(3) hadronic  mean-field model is written as \cite{Papazoglou:1998vr}

\begin{align}
	\mathcal{L}_{\text{chiral}} &= \mathcal{L}_{\text{kin}} + \sum_{M=P,X,V,A} \mathcal{L}_{\text{BM}} +
	\mathcal{L}_{0}+ \mathcal{L}_{\text{vec}} + \mathcal{L}_{\text{SB}}. \label{EQ1}
\end{align}

%Eq (\ref{EQ1}) represents a relativistic quantum field theoretical model of mesons and baryons, by incorporating a nonlinear realization of chiral symmetry \cite{weinberg1968nonlinear,coleman1969structure,bardeen1969some} and broken scale invariance \cite{papazoglou1999nuclei,mishra2004effects,zschiesche2004medium} to describe strongly interacting nuclear matter.

% The medium-modified properties of baryons and mesons in hot and dense hadronic matter are extensively studied using this model. 
In Eq. (\ref{EQ1}), \(\mathcal{L}_{\text{kin}}\) represents the kinetic energy term for baryons and mesons
and is given by
%  ,the term \(\mathcal{L}_{\text{BM}}\) represent the baryon-meson interactions where the interaction terms involving baryon and spin-0 mesons contribute to the generation of baryon masses. \(\mathcal{L}_{\text{vec}}\) includes quartic self-interactions term and reproduces the dynamical mass generation of the vector mesons through interaction with scalar mesons. \(\mathcal{L}_{\text{0}}\) represents the spontaneous breaking of chiral symmetry of meson-meson interaction terms. The explicit breaking of chiral symmetry is described by \(\mathcal{L}_{\text{SB}}\). 
%The kinetic energy terms are given as
\begin{align}
	\mathcal{L}_{\text {kin }}&=i \operatorname{Tr} \bar{B} \gamma_{\mu} D^{\mu} B+\frac{1}{2} \operatorname{Tr} D_{\mu} X D^{\mu} X + \operatorname{Tr}\left(u_{\mu} X u^{\mu} X + X u_{\mu} u^{\mu} X\right)
	\quad+\frac{1}{2} \operatorname{Tr} D_{\mu} Y D^{\mu} Y \nonumber \\
	+&\frac{1}{2} D_{\mu} \chi D^{\mu} \chi-\frac{1}{4} \operatorname{Tr}\left(V_{\mu \nu} V^{\mu \nu}\right)-\frac{1}{4} \operatorname{Tr}\left(\mathcal{A}_{\mu \nu} \mathcal{A}^{\mu \nu}\right).  
	\label{Eq_kinetic_2}
\end{align}
In Eq. (\ref{Eq_kinetic_2}), the first term represents the kinetic energy term for the baryon octet, $B$.  
The covariant derivative $D_{\mu}$ appearing in this term is defined as $D_\mu B = \partial_\mu B + i \left[\Gamma_\mu, B\right]$, with
$\Gamma_{\mu}=-\frac{i}{4}\left[u^{\dagger} \partial_{\mu} u-\partial_{\mu} u^{\dagger} u+\right.$ $\left.u \partial_{\mu} u^{\dagger}-\partial_{\mu} u u^{\dagger}\right]$.
Here, $u=\exp \left[\frac{i}{\sigma_{0}} \pi^{a} \lambda^{a} \gamma_{5}\right]$ is the unitary transformation operator through which pseudoscalar mesons enter into the calculations.
%As will be presented later in detail, this kinetic term defined in terms of covariant derivative   contributes to the interactions of kaons and antiakons with the nucleons and hyperons in the strange medium (popularly known as Weinberg Tomozawa term). 
The second and third terms of Eq. (\ref{Eq_kinetic_2})  are the kinetic terms for the scalar and pseudoscalar  mesons. Fourth and fifth terms represent the kinetic terms for the pseudoscalar singlet $Y$ and the dilaton field $\chi$, respectively. The last two terms  defined in terms of field tensors, $ V_{\mu \nu}$ and
$ \mathcal{A}_{\mu \nu}$, represent the  kinetic term of spin-1 vector and axial vector mesons, respectively.

%In the above equation, \( P \) represents the pseudoscalar singlet, and \( B(S) \) denotes the multiplet of baryons (scalar) mesons. Also, \( V_{\mu \nu}=D_{\mu} V_{\nu}-D_{\nu} V_{\mu} \) is the vector meson field tensor and \( A_{\mu \nu}=D_{\mu} A_{\nu}-D_{\nu} A_{\mu} \) (\( \mathcal{A}_{\mu \nu}=\partial_{\mu} \mathcal{A}_{\nu}-\partial_{\nu} \mathcal{A}_{\mu} \)),  is the axial-vector field tensor. $F_{\mu \nu}$ is the field strength tensor of the photon. Further, in the fifth term (glueball or dilaton field) of the above equation, there is no change if we use covariant derivatives instead of ordinary derivatives because the additional commutator term gives zero contribution. \( P \) is independent of the octet and therefore has its own kinetic term. In the above, $u_{\mu}=-\frac{i}{2}\left[u^{\dagger} \partial_{\mu} u-u \partial_{\mu} u^{\dagger}\right]$, where $u=\exp \left[\frac{i}{\sigma_{0}} \pi^{a} \lambda^{a} \gamma_{5}\right]$ is the unitary transformation operator. Field $\Phi \equiv B, X, Y, A_{\mu}, V_{\mu}$ has a covariant derivative that is expressed as follows: 
%$D_{\mu} \Phi=\partial_{\mu} \Phi+\left[\Gamma_{\mu}, \Phi\right]$ and $\Gamma_{\mu}=-\frac{i}{4}\left[u^{\dagger} \partial_{\mu} u-\partial_{\mu} u^{\dagger} u+\right.$ $\left.u \partial_{\mu} u^{\dagger}-\partial_{\mu} u u^{\dagger}\right]$. \\

Within the chiral SU(3) model, to describe the interactions of nucleons through the exchange of scalar fields $\sigma, \zeta$ and $\delta$ and vector fields $\omega$ and $\rho$, under the
mean-field approximation,  following interaction Lagrangian is written
\begin{align}
	\mathcal{L}_{B M}  =-\sum_{i} \overline{\psi_{i}}\left[g_{ \omega i} \gamma_{0} \omega+g_{ \rho i} \gamma_{0} \tau_{3} \rho+   m_{i}^{*} \right] \psi_{i}, \label{Eq_BM_intern}
\end{align}
where $i = p,n$.
In the above, $m_i^{*}$ is the effective mass of nucleons defined as
\begin{align}
	m_i^*=-\left(g_{\sigma i} \sigma+g_{\zeta i} \zeta+g_{\delta i} \tau_3 \delta\right) \label{Eq_of_baryon_mass}.
\end{align}
The coupling constants \( g_{\sigma i} \), \( g_{\zeta i} \) and \( g_{\delta i} \)
describe the strength of interactions of nucleons with scalar fields.
The term $\mathcal{L}_{\text {0}}$
of Eq. (\ref{EQ1}) gives the self interaction of scalar mesons
$\sigma, \zeta$ and $\delta$ and also the interactions for dilaton field $\chi$. This term is written as
\begin{align}
	\mathcal{L}_{\text {0}} & = 
	- \frac{ 1 }{ 2 } k_0 \chi^2
	(\sigma^2+\zeta^2+\delta^2) + k_1 (\sigma^2+\zeta^2+\delta^2)^2
+ k_2 ( \frac{ \sigma^4}{ 2 } + \frac{\delta^4}{2} + \zeta^4
	 +3 \sigma^2 \delta^2)
	  	\nonumber \\
	  	     &   + k_3 \chi (\sigma^2 - \delta^2) \zeta 
	-k_{4} \chi^{4}-\frac{1}{4} \chi^{4} \ln \frac{\chi^{4}}{\chi_{0}^{4}}+\frac{\delta}{3} \chi^{4} \ln \frac{\left(\sigma^{2}-\delta^{2}\right) \zeta}{\sigma_{0}^{2} \zeta_{0}}.
\label{Eq_meson_self_scale1}
\end{align}
 In Eq. (\ref{Eq_meson_self_scale1}), last two terms account for scale breaking effects introduced in the chiral SU(3) model through the dilation field $\chi$.
In  Eq. (\ref{EQ1}), the term $\mathcal{L}_{\text {vec }}$  describes the self interactions of vector mesons through the Lagrangian density
\begin{align}
	\mathcal{L}_{\text {vec }} & = \frac{1}{2} \frac{\chi^2}{\chi_0^2}\Big(
	m_{\omega}^{2} \omega^ 2+m_{\rho}^{2} \rho^ 2
	\Big) +g_4 (\omega^4 +6 \omega^2 \rho^2+\rho^4). 
	\label{Eq_vec_lag1} 
\end{align}
The  term, $\mathcal{L}_{S B}$, of Eq. (\ref{EQ1})
is the explicit symmetry breaking term and is given by
\begin{align}
\mathcal{L}_{S B} & =- \left(\frac{\chi}{\chi_0}\right)^2\left[m_{\pi}^{2} f_{\pi} \sigma+\left(\sqrt{2} m_{K}^{2} f_{K}-\frac{1}{\sqrt{2}} m_{\pi}^{2} f_{\pi}\right) \zeta\right].
\end{align}

At zero temperature, the thermodynamic potential per unit volume of the grand canonical ensemble is expressed as 
\begin{equation}
	\begin{array}{r}
	\frac{\Omega}{V}= -\sum_i \frac{\gamma_i  }{(2 \pi)^3} \int_0^{k_{F,i}} d^3 \mathbf{k}\left[E_i^*(\mathbf{k})-\mu_i^*\right]
	-\mathcal{L}_0  
		 -\mathcal{L}_{\text {vec }}-\mathcal{L}_{\mathrm{SB}}-\mathcal{V}_{v a c}, \label{Thermodynaics_potn}
	\end{array}
\end{equation}
where $E_i^*(\mathbf{k}) = \sqrt{\mathbf{k}^2 + m_i^{*2}}$ and $\mu_i^* = \mu_i  - g_{\omega i} \omega-g_{\rho i} \tau_3 \rho$ .
% are the single-particle effective energy and effective chemical potential, respectively, at temperature $T$. The spin degeneracy factor is represented by $\gamma_i$ and factor, $\beta = \frac{1}{kT}$.
  In addition, the vacuum potential energy, $\mathcal{V}_{\text{vac}}$, is subtracted from Eq. (\ref{Thermodynaics_potn}) to achieve vanishing vacuum energy.
The equations of motion for the non-strange $\sigma$, the strange scalar $\zeta$, the scalar isovector $\delta$, the vector $\omega$, the vector-isovector $\rho$,  and the scalar dilaton $\chi$ fields are derived by minimizing the thermodynamic potential and are written as

\begin{align}
	& k_{0} \chi^{2} \sigma-2 k_{2}\left(\sigma^{3}+3 \sigma \delta^{2}\right) -2 k_{3} \sigma \zeta \chi -4 k_{1}\left(\sigma^{2} +\delta^{2} +\zeta^{2}\right) \sigma \nonumber \\ 
	& -\frac{d}{3} \chi^{4}\left(\frac{2 \sigma}{\sigma^{2}-\delta^{2}}\right)+\left(\frac{\chi}{\chi_{0}}\right)^{2} m_{\pi}^{2} f_{\pi}-\sum_{i} g_{\sigma i} \rho_{i}^{s}=0, \label{eq_sigma} \\
	&  k_{0} \chi^{2} \zeta-4 k_{2} \zeta^{3}+k_{3} \chi\left(\delta^{2}-\sigma^{2}\right) - 4 k_{1}\left(\sigma^{2}+\delta^{2}+\zeta^{2}\right) \zeta \nonumber \\
	& -\frac{d}{3} \frac{\chi^{4}}{\zeta}+\left(\frac{\chi}{\chi_{0}}\right)^{2}\left[\sqrt{2} m_{K}^{2} f_{K}-\frac{1}{\sqrt{2}} m_{\pi}^{2} f_{\pi}\right]-\sum_{i} g_{\zeta i} \rho_{i}^{s}=0, \label{eq_zeta}
\end{align}
\begin{align}
	& k_{0} \chi^{2} \delta +2 k_{3} \chi \delta \zeta -2 k_{2}\left(\delta^{3}+3 \sigma^{2} \delta\right)-4 k_{1}\left(\sigma^{2} +\delta^{2} +\zeta^{2}\right) \delta \nonumber \\
	& +\frac{2}{3} d \chi^{4}\left(\frac{\delta}{\sigma^{2}-\delta^{2}}\right)-\sum_{i} g_{\delta i} \tau_{3} \rho_{i}^{s}=0,
	\label{eq_delta}\\
	&\left(\frac{\chi}{\chi_{0}}\right)^{2} m_{\omega}^{2} \omega+g_{4}\left(12 \rho^{2} \omega + 4 \omega^{3}\right)-\sum_{i} g_{\omega i} \rho_{i}=0, \label{eq_omega} \\
	&\left(\frac{\chi}{\chi_{0}}\right)^{2} m_{\rho}^{2} \rho+g_{4}\left( 12 \omega^{2} \rho + 4 \rho^{3} \right)-\sum_{i} g_{\rho i} \tau_{3} \rho_{i}=0, \label{eq_rho} 
%	\frac{\partial(\Omega / V)}{\partial \phi}=\left(\frac{\chi}{\chi_{0}}\right)^{2} m_{\phi}^{2} \phi+8 g_{4} \phi^{3}-\sum_{i} g_{\phi i} \rho_{i}=0, \label{eq_phi}
\end{align}
and
\begin{align}
	&k_{0} \chi\left(\sigma^{2}+\delta^{2}+\zeta^{2}\right)+\left(4 k_{4}-d\right) \chi^{3}+k_{3}\left(\delta^{2}-\sigma^{2}\right) \zeta+\chi^{3}\left[1+\ln \left(\frac{\chi^{4}}{\chi_{0}^{4}}\right)\right] \nonumber \\
	&-\frac{4}{3} d \chi^{3} \ln \left(\left(\frac{\left(\sigma^{2}-\delta^{2}\right) \zeta}{\sigma_{0}^{2} \zeta_{0}}\right)\left(\frac{\chi}{\chi_{0}}\right)^{3}\right)+\frac{2 \chi}{\chi_{0}^{2}}\left[m_{\pi}^{2} f_{\pi} \sigma+\left(\sqrt{2} m_{K}^{2} f_{K}-\frac{1}{\sqrt{2}} m_{\pi}^{2} f_{\pi}\right) \zeta\right] \nonumber \\
	&-\frac{\chi}{\chi_{0}^{2}}\left(m_{\omega}^{2} \omega^{2}+m_{\rho}^{2} \rho^{2}  \right)=0, 
	\label{eq_chi}
\end{align}
respectively.
%respectively. The scalar and vector densities of the nucleons appearing in the above equation are defined as
%\begin{align}
%	\rho_i^s=\gamma_i \int \frac{d^3 k}{(2 \pi)^3} \frac{m_i^*}{E_i^*(\mathbf{k})}\left(n_i(\beta)+\bar{n}_i(\beta)\right),
%	\label{Eq_scalar_d}
%\end{align}
%and
%\begin{align}
%	\rho_i^v=\gamma_i \int \frac{d^3 k}{(2 \pi)^3}\left(n_i(\beta)-\bar{n}_i(\beta)\right).
%	\label{Eq_vector_d}
%\end{align}
%with
%\begin{align}
%	n_i(\beta)=\frac{1}{\exp \left[\beta\left(E_i^*(\mathbf{k})-\mu_i^*\right) \right]+1},   \quad \bar{n}_i(\beta)=\frac{1}{\exp \left[\beta\left(E_i^*(\mathbf{k})+\mu_i^*\right)\right] + 1}, \label{Eq_distribution_fun}
%\end{align}
%as the finite temperature distribution function of baryons and antibaryons, respectively.
%In the present work, we shall investigate the in-medium properties of pseudoscalar and vector mesons at finite isospin asymmetry and finite strangeness fraction of the medium. The isospin asymmetry is introduced through the parameter $\eta = \frac{-\sum_i I_{3i} \rho_i^v}{ \rho_B}$, whereas for the finite strangeness fraction the definition  $f_s = \frac{\sum_i |s_i| \rho_i^v}{\rho_B}$
%is used. The symbols $I_{3i}$, $|s_i|$, and $\rho_B$ represent the isospin quantum number (third component), the number of strange quarks, and the total baryonic density, respectively.
In the present work, we shall discuss the formation of bound states for the nuclei  $^{12}$C, $^{16}$O, $^{40}$Ca
and $^{208}$Pb. 
As we shall describe later, the optical potentials of mesons
required as input in the Klein Gordan equation for the study of bound states, will depend upon the scalar fields $\sigma, \zeta$ and $\delta$.
For a nuclei with radius $R$, the values of scalar fields are required  as a function of radial coordinator $r$, such that $r$ varies from zero to $R$.
 For this the coupled equations of motion for scalar and vector fields are solved for the baryon densities corresponding to these nuclei as a function of $r$. In the present work, we have considered the radial dependence of vector densities of nucleons  
  through the parameterization in the form of harmonic oscillator or two parameter Fermi distribution function, as discussed below, and the second order derivative term of the form $-\nabla^2 {\cal \sigma}(r)$ in the field equations given above is not considered.
 
Total baryon density, $\rho_B$, in a given nuclei is sum of densities of protons and neutrons, i.e.,
 \begin{align}
\rho_B (r) = \sum_{i = p,n} \rho_i (r).
\label{Eq_den_dis1}
 \end{align}
For the  nuclei upto $^{18}O$, we shall use the harmonic oscillator type density distribution, given as \cite{Nieves:1993ev}
\begin{align}
\rho_i (r) = \rho_{i,0} \left(1 + a_i \left(\frac{r}{R_i}\right)^2\right) \exp \left[-\left(\frac{r}{R_i}\right)^2\right].
\label{Eq_density_dis_HO}
\end{align}
   On the other side, for heavy nuclei, two parameter Fermi distribution function will be used in the present work \cite{Nieves:1993ev} and is written as 
\begin{align}
\rho_i (r) = \frac{\rho_{i,0}}{1 + \exp \left[\frac{(r-R_i)}{a_i}\right]}.  
\label{Eq_density_dis_Fermi}
\end{align}   
In Eqs. (\ref{Eq_density_dis_HO}) and  (\ref{Eq_density_dis_Fermi}), $\rho_{i,0}$ is the density of nucleons at the center of nuclei and   $R_i$  and $a_i$  are the radii and diffuseness  parameters, respectively corresponding to nucleons of type $i$. Values of these parameters for different nuclei are tabulated in Table I of Ref.  \cite{Nieves:1993ev}.
For the nuclei with mass number $A$ and  having different number of protons, $Z$ and neutrons, $N$, the isospin asymmetry parameter
$I$ is given by, $I = (N-Z)/A$, which is further defined in terms of densities of nucleons as, $I = \frac{\rho_n(r) - \rho_p(r)}{2\rho_B(r)}$.
The system of equations  given by Eqs. (\ref{eq_sigma}) to (\ref{eq_chi}) are solved for the densities in a given nuclei from center to surface, as given by Eq. (\ref{Eq_den_dis1}).
At zero temperature, the number density $\rho_i$ and the scalar density $\rho_i^s$ are related to the Fermi momentum $k_{F,i}$ through the relations
\begin{align}
\rho_{i} = \frac{\gamma_ik_{F,i}^3}{6 \pi^2},
\label{Eq_den_v_kf}
\end{align}
and 
\begin{align}
\rho_i^s=\gamma_i  \int_0^{ k_{F, i}} \frac{d^3 \mathbf{k}}{(2 \pi)^3} \frac{m_i^*}{E_i^*(\mathbf{k})}=\frac{\gamma_i  m_i^*}{4 \pi^2}\left[k_{F, i} E_i^*-m_i^{* 2} \ln \left(\frac{k_{F, i}+E_i^*}{m_i^*}\right)\right] ,
\label{Eq_salar_dens1}
\end{align}
respectively. 
In the above, $\gamma_i$ is the degeneracy factor for the nucleons.
To study the properties of mesons in infinite nuclear medium at zero temperature,  
 Eqs. (\ref{eq_sigma}) to (\ref{eq_chi})
 are solved using the expressions of
  $\rho_i$ and $\rho_i^s$ given above.
  For finite temperature effects, 
  $\rho_i$ and $\rho_i^s$ are defined as
 \begin{eqnarray}
 \rho_{i} = \gamma_{i}\int\frac{d^{3}\mathbf{k}}{(2\pi)^{3}}  
 \Bigg(\frac{1}{1+\exp\left[\beta(E^{\ast}_i(\mathbf{k}) 
 -\mu^{*}_{i}) \right]}-\frac{1}{1+\exp\left[\beta(E^{\ast}_i(\mathbf{k})
 +\mu^{*}_{i}) \right]}\Bigg) ,
 \label{rhov0}
 \end{eqnarray}
 
  and
 
 \begin{eqnarray}
 \rho_{i}^{s} = \gamma_{i}\int\frac{d^{3}\mathbf{k}}{(2\pi)^{3}} 
 \frac{m_{i}^{*}}{E^{\ast}_i(\mathbf{k})} \Bigg(\frac{1}{1+\exp\left[\beta(E^{\ast}_i(\mathbf{k}) 
 -\mu^{*}_{i}) \right]}+\frac{1}{1+\exp\left[\beta(E^{\ast}_i(\mathbf{k})
 +\mu^{*}_{i}) \right]}\Bigg),
 \label{rhos0}
 \end{eqnarray}
 respectively. Here, $\beta = 1/T$.
\section{Optical potentials of mesons}
\label{sec:potential}
To obtain the binding energy and absorption decay width of the mesic nuclei, we shall solve the Klein Gordon equation in the presence of local potential $V(r)$ and is written as
\begin{align}
\left(-\nabla^2 + \left(\mu +  V(r)\right)^2\right)\psi(r) = \epsilon^2 \psi(r).
\label{Eq_KG_Coord1}
\end{align}  
In the above equation, $\mu$ is the reduced mass of meson and nuclei under study. The local potential $V(r)$ is complex in nature, i.e., 
\begin{align}
V(r) = U(r) - \frac{i}{2} W(r),
\end{align}
where the real part $U(r)$ is related to the mass shift of mesons, whereas, the imaginary part $W(r)$ relates to the absorption of mesons in the nuclei under consideration. For the real part, $U(r)$, we write \cite{Metag:2017yuh}
\begin{align}
U(r) = m_\psi^*(r)-m_\psi = \Delta m_\psi (\rho_0) \frac{\rho_B(r)}{\rho_0},
\label{Eq_real_opt1}
\end{align}
where  $\rho_0$ denotes the baryon density at the center of nuclei and  $\Delta m_\psi (\rho_0)$ $[\psi = \eta, \eta^{'}, D^{0}, \bar{D^0}, B^{0},$ $ \bar{B^0},  \bar{K^0}, \phi ]$ is the mass shift at this density.
Also, $\rho_B(r)$ is the baryon density distribution inside the nuclei and is given by Eq.(\ref{Eq_den_dis1}).
The potential $W(r)$  is related to the 
decay width $\Gamma_0$ through the relation \cite{Metag:2017yuh}
\begin{align}
W(r) = \Gamma_0 (\rho_0)\frac{\rho_B(r)}{\rho_0}.
\label{Eq_imag_decay_W1}
\end{align}
In the present work,   the value of  $\Gamma_0 (\rho_0)$ for pseudoscalar mesons, $\eta, \eta^{'}, D^{0}, \bar{D^0}, B^{0}, \bar{B^0}$ and  $\bar{K^0}$ will be calculated following the procedure adopted in the calculations within QMC model \cite{Cobos-Martinez:2023hbp}. Under this,
$\Gamma_0 (\rho_0)$ is related to the mass shift $\Delta m_\psi (\rho_0)$ through a parameter $\kappa$ using relation
\begin{align}
\Gamma_0 (\rho_0) = - \kappa \Delta m_\psi (\rho_0) + \Gamma_{\text{vac}}.
\label{Eq_imga_decay1}
\end{align}
The parameter $\kappa$  imitate the absorption of mesons in the nuclear medium and $\Gamma_{\text{vac}}$ is the decay width in the vacuum.
In our present work, the contributions from the first term of above equation will be considered
to estimate the contribution of imaginary part of potential to the binding energy.
 We shall solve the Klein Gordon equation in the momentum space using the Fourier transformation and following the partial wave decomposition method as described in Refs. \cite{Kwan:1978zh,Landau:1982iu} to obtain the complex energy eigen values $\epsilon$ which are further related to the binding energy $\cal E_B$ and decay width $\Gamma$ through relations $\cal E_B = \text{Re}\,\epsilon - \mu$ and $\Gamma = - 2\, \text{Im}\, \epsilon$, respectively. 
 In the following, we briefly describe the calculations of mass shift of $\eta, \eta^{'}, D^{0}, \bar{D^0}, B^{0}, \bar{B^0}$,  $\bar{K^0}$
 and $\phi$ mesons which will be used as input to obtain the optical potentials of mesons in the environment of different nuclei. 
\subsection{Pseudoscalar $\eta$ and $\eta^{'}$ mesons}
In this section, we present the details of interaction Lagrangian density and dispersion relations to obtain the in-medium masses of pseudoscalar $\eta$ and $\eta^{\prime}$ mesons in the nuclear medium relevant for different nuclei. The physical states $\eta$ and $\eta^{\prime}$ result due to the mixing of $\eta_8$ and $\eta_0$ states. The pseudoscalar $\eta_8$ appears in the octet of pseudoscalar mesons defined by \cite{Papazoglou:1998vr}
\begin{equation}
P = \frac{1}{\sqrt{2}} \pi_a \lambda^a = \left(\begin{array}{ccc}\frac{1}{\sqrt{2}}\left(\pi^0+\frac{\eta_8}{\sqrt{1+2 w^2}}\right) & \pi^{+} & 2 \frac{K^{+}}{w+1} \\ \pi^{-} & \frac{1}{\sqrt{2}}\left(-\pi^0+\frac{\eta_8}{\sqrt{1+2 w^2}}\right) & 2 \frac{K^0}{w+1} \\ 2 \frac{K^{-}}{w+1} & 2 \frac{K^0}{w+1} & -\frac{\eta_8 \sqrt{2}}{\sqrt{1+2 w^2}}\end{array}\right),
\end{equation}
where  $w = \sqrt{2}\sigma_{0}/\zeta_{0}$. For the case of pseudoscalar single $\eta_0$, we define
\begin{equation}
Y = \frac{1}{\sqrt{3}} \eta_0 I.
\label{eq_singlet_def1}
\end{equation}
  
Within the chiral SU(3) model, for the pseudoscalar $\eta_8$, the interaction   Lagrangian density is written as
	\begin{eqnarray} \label{etaN}
		\mathcal{L}_{{\eta_{8}} B}  &=&
			-\frac{1}{2}\left(
				m_{\eta_8}^2
				-\frac{(\sqrt{2}\sigma ^\prime -4 \zeta ^\prime )m^2_\pi f_\pi + 8 \zeta ^\prime m^2_K f_K}{\sqrt{2} f^2}
				\right) \eta_8^2\nonumber\\
	&+&	\left( \frac{1}{2}-\frac{\sqrt{2}\sigma ^\prime f_\pi + 4 \zeta ^\prime (2 f_K-f_\pi) }{\sqrt{2}f^2} \right) \partial^{\mu}\eta_8\partial_{\mu}\eta_8 \nonumber\\
		&&+\frac{d^{'}}{4 f^{2}}\left( \rho_{p}^{s}+\rho_{n}^{s}    \right)\partial^{\mu}\eta_8\partial_{\mu}\eta_8,
	\label{Eq_inter_eta8}
	\end{eqnarray}
	where $d'$ = $3d_1+d_2$. Also, $\sigma^{'}$ and $\zeta^{'}$ are the
	fluctuations from the vacuum expectation values, i.e., 
	$\sigma^{'} = \sigma - \sigma_0$ and $\zeta^{'} = \zeta - \zeta_{0}$.
	The first term of Eq.(\ref{Eq_inter_eta8}) (mass term) corresponds to the explicit symmetry breaking  and is obtained from the general term 
	\begin{equation}
	 \label{esb-gl}
	 \mathcal{L_{\text{Mass term}}} =
	 -\frac{1}{2} \Tr A_p \left(uXu+u^{\dagger}Xu^{\dagger}\right).
	\end{equation}	
In the above equation, $A_p$ is a diagonal matrix given as
	 \begin{equation}
	\label{apmat}
	 		A_p=\frac{1}{\sqrt{2}}
	 		\left( \begin{array}{ccc}
	 			m_{\pi}^2 f_{\pi}& 0& 0\\   
	 			0 & m_\pi^2 f_\pi& 0\\
	 			0 & 0& 2 m_K^2 f_K
	 			-m_{\pi}^2 f_\pi
	 		\end{array} \right).
	 	\end{equation}

The vacuum mass $m_{\eta_8}$ of $\eta_8$ meson deduced from Eq.(\ref{esb-gl}) is identified as
	 	\begin{equation}
	 		m_{\eta_8}=\frac{1}{f}\sqrt{\frac{1}{2}(-\left(8 f_K f_\pi (m_\pi^2+m_K^2))+16 f_K^2 m_K^2 +6 f_\pi^2 m_\pi^2\right)},
	 	\end{equation}
	  where $f = \sqrt{f_\pi^2 +2 \left(2f_K - f_\pi\right)^2}$.
The second term of Eq.(\ref{Eq_inter_eta8}) is obtained from the general kinetic term of the pseudoscalar mesons (third term of  Eq. (\ref{Eq_kinetic_2})) and is also knwon as first range term in the chiral SU(3) model.

The last term of Eq.(\ref{Eq_inter_eta8}), in terms of $d_1$ and $d_2$, is  obtained from the Lagrangian densities (known as $d_1$ and $d_2$ terms)
 \cite{Mishra:2006wy,Mishra:2008kg}	
 	\begin{equation}
 		{\cal L }_{d_1}^{\bar{B}B} =\frac {d_1}{2} \Tr (u_\mu u ^\mu) \Tr(\bar{B} B), \quad\quad \quad \quad
 		{\cal L }_{d_2}^{\bar{B} B} =d_2 \Tr (\bar {B} u_\mu u ^\mu B).
 		\label{Eq_d1d2_L2}
 	\end{equation}
 In the present work for $\eta_8$ mesons, from the  Weinberg Tomozawa term  we do not obtain interaction terms with finite contributions unlike the case for kaons \cite{Mishra:2006wy,Mishra:2008kg} and $D$ \cite{Mishra:2008cd}  mesons studied in past using the present chiral SU(3) mean field model. These observations are also consistent with the calculations of $\eta$ mesons properties in the nuclear medium at zero temperature using chiral perturbation theory \cite{Zhong:2006jj}.

%%%%%%%%%%%%%%%%%%%%%%%%%%%55	
From the  Lagrangian density given in Eq.(\ref{Eq_inter_eta8}), the equations of motion for $\eta_8$ is obtained, whose Fourier transformation leads to the  dispersion relation
%\begin{eqnarray} 
%		&& \partial^{\mu}\partial_{\mu} \eta + \left(
%		m_{\eta}^2-\frac{(\sqrt{2}\sigma ^\prime -4 \zeta ^\prime )m^2_\pi f_\pi + 8 \zeta ^\prime m^2_K f_K}{\sqrt{2} f^2}
%		\right)\eta  \nonumber\\
%		&&+\frac{2}{f^{2}}\left(\frac{d^{\prime}  \Sigma \rho_{i}^{s}}{4}+\frac{d_{2} \rho_{\Lambda^{0}}^{s}}{2}+\frac{3d_{2} (\rho_{\Xi^-}^{s}+\rho_{\Xi^0}^{s})}{4}-\frac{\sqrt{2} \sigma^{\prime} f_{\pi}+4 \zeta^{\prime}\left(2 f_{K}-f_{\pi}\right)}{\sqrt{2}}\right) \partial^{\mu} \partial_{\mu} \eta=0.
%		\label{eom}
%	\end{eqnarray}
%% In Eq.(\ref{eom}), $\rho_{b}^{s}$ represents the net scalar baryon density and  $d'$ = $3d_1+d_2$ \cite{Rajesh2020eta}. 
%The dispersion relation for $\eta$ meson can be obtained by the Fourier transform of the above equation, $i.e.$,
	\begin{equation}
		-\omega^2+ { \textbf{k}}^2 + m_{\eta_8}^2 + \Pi_{\eta_8}(\omega, | \textbf{k}|, \rho_i(r))=0,
		\label{drk}
	\end{equation}	
	where  $\Pi_{\eta_8}$ stands for $\eta_8$ meson's   self-energy and is given  as,
	\begin{eqnarray}
		\Pi_{\eta_8} (\omega, | \textbf{k}|, \rho(r)) &= &
		- \frac{ 8 \zeta ^\prime m^2_K f_K + (\sqrt{2}\sigma ^\prime -4 \zeta ^\prime )m^2_\pi f_\pi}{\sqrt{2} f^2}
	+	\left[
		 \frac{d^{\prime}}{2f^{2}}  \left(\rho_{p}^{s} + \rho_{n}^{s}\right) \right.  \nonumber\\
		&-& \frac{2}{f^2} \left( \left. \frac{\sqrt{2} \sigma ^\prime f_\pi + 4 \zeta ^\prime (2 f_K-f_\pi) }{\sqrt{2}} \right)\right]
		(-\omega ^2 + {\textbf{ k}}^2). 
		\label{sen}
	\end{eqnarray}
At zero momentum ($\textbf{k} = 0$), the effective mass  $m_{\eta_8}^{*}$ is written as
\begin{equation}
m_{\eta_8}^{*} = \sqrt{m_{\eta_8}^{2} + 		\Pi_{\eta_8} (\omega, | \textbf{k}|, \rho_i(r))}.
\end{equation}
For the pseudoscalar singlet $\eta_0$, within the chiral SU(3) hadronic mean field model, we write
\begin{equation}
	\mathcal{L} = \frac{1}{2} \text{Tr} D_\mu Y D^\mu  Y - \frac{1}{2} m_{\eta_0}^2\, \text{Tr} Y^2 - \frac{
	1}{2} \text{Tr} A_p \left( u(X+iY)u + u^{\dagger}(X-iY)u^{\dagger}\right).
\label{eq_eta0_iter1}
\end{equation}
In above, the first term is the is the kinetic energy term for 
pseudoscalar mesons. The second term is the part of explicit symmetry breaking term and gives the finite mass to the pseudoscalar singlet. The last term of Eq.(\ref{eq_eta0_iter1})
is the extension of Eq.(\ref{esb-gl}) where now the singlet $Y$ is included.  This leads to the terms having mixing between $\eta_0$ and $\eta_8$ states. Explicitly, using Eq.(\ref{eq_singlet_def1}) in
Eq.(\ref{eq_eta0_iter1}), we get 
\begin{equation}
\mathcal{L} = \frac{1}{2} \partial_\mu \eta_0  \partial^\mu \eta_0 - \frac{1}{2} m_{\eta_0}^2 \eta_0^2 + \frac{1}{\sqrt{3} \sigma_{0} c} \left(m_\pi^2 f_\pi - m_K^2 f_K\right) \eta_0 \eta_8,
\label{Eq_eta0eta8_int1}
\end{equation}
where, $c= {2 \sigma_0}/{(\sqrt{2} \zeta_0 + \sigma_0)}$.
In the last term of above equation the interaction terms corresponding to pure $\eta_8$ states, which are given in Eq.(\ref{Eq_inter_eta8}), are not repeated to avoid double counting. From Eq.(\ref{Eq_eta0eta8_int1}) we observe that unlike $\eta_8$, for pure $\eta_0$ state contributions from the in-medium interactions does not appear, i.e., 
\begin{equation}
m_{\eta_0}^* = m_{\eta_0}.
\label{Eq_meta0_med1}
\end{equation}
From the mixed term (last term of Eq.(\ref{Eq_eta0eta8_int1})), we identify
\begin{equation}
m_{\eta_0 \eta_8}^{*} = 
  \left[ \frac{2}{\sqrt{3} \sigma_0 c} 
\left(m_\pi^2 f_\pi - m_K^2 f_K\right) \right]^{1/2}.
\label{Eq_meta0eta8med1}
\end{equation}
The effective masses $m_\eta^{*}$ and $m_{\eta^{'}}^{*}$  of physical $\eta$ and $\eta^{'}$ mesons are obtained by diagonalization of mass matrix 
\begin{equation}
m_{ij} = 
\begin{pmatrix}
m_{\eta_0}^* & m_{\eta_0\eta_8}^* \\
m_{\eta_8\eta_0}^* & m_{\eta_8}^*
\end{pmatrix}
.
\label{Eq_mass_matrix_eta}
\end{equation}

%previous lines The parameter $d'$ is fitted to the values of scattering length, $a^{\eta N}$ \cite{Zhong2006}. For the present model, the scattering length expression is obtained as
% \textbf{The parameter $d'$ is evaluated from the experimental values of the scattering length $a^{\eta N}$.

 The  parameter $d^{'}$ is expressed in terms of $\eta-N$ scattering length, $a^{\eta N}$. The expression for scattering length $a^{\eta N}$  calculated within chiral SU(3) model is given by
\cite{Kumar:2020gpu}  %question11
\begin{eqnarray}
		a^{\eta N} &=& \frac{1}{4 \pi \left (1+\frac{m_\eta}{M_N}\right )} \Big [ \Big( \frac{d'}{\sqrt{2}}-\frac{g_{\sigma N}{f_\pi}}{m^2_\sigma}+\frac{4 (2f_K-f_\pi) g_{\zeta N}}{m^2_\zeta} \Big) \frac {m_\eta ^2} {\sqrt{2}f^2} \nonumber \\
		&+& \left( \frac{\sqrt{2} g_{\sigma N}}{m^2_\sigma}-\frac{4 g_{\zeta N}}{m^2_\zeta} \right )\frac {m^2_\pi f_\pi} {2\sqrt{2}f^2}+ \frac{2\sqrt{2} g_{\delta N}}{m^2_\delta} \frac {m^2_K f_K} {f^2}  \Big ],
		\label{sl}
	\end{eqnarray}
% After rearranging for $d'$, the expression can be written in terms of $a^{\eta N}$ and other model parameters, i.e.,
% \begin{eqnarray}
%		d' &=& 8 \pi {\left (1+\frac{m_\eta}{M_N}\right )} f^2 \frac{ a^{\eta N}}{m^2_\eta} +\frac{\sqrt{2}g_{\sigma N}}{m^2_\sigma}-\frac{4 \sqrt{2} (2f_K-f_\pi) g_{\zeta N}}{m^2_\zeta}  \nonumber \\
%		&-& \left( \frac{\sqrt{2} g_{\sigma N}}{m^2_\sigma}-\frac{4 g_{\zeta N}}{m^2_\zeta} \right )\frac {m^2_\pi f_\pi} {\sqrt{2} m^2_\eta}- \frac{4\sqrt{2} g_{\delta N}m^2_K f_K}{m^2_\delta m^2_\eta}.
%		\label{dp}
%	\end{eqnarray} 
 which can be rearranged to obtain the expression for  $d^{'}$. We   calculate $d'$ for the value of $a^{\eta N} =$  0.91   \cite{Zhong:2006jj}.
 From the above calculated effective masses $m_\eta^*$ and $m_{\eta^{\prime}}^*$ of $\eta$ and $\eta^{\prime}$ mesons, corresponding mass shift will be calculated which shall be used as input in Eqs. (\ref{Eq_real_opt1}) and (\ref{Eq_imga_decay1})
 to obtain the optical potentials of these mesons in the nuclei.

\subsection{Pseudoscalar $D^{0}$ and $\bar{D^{0}}$ mesons}
The effective masses of neutral
$D^{0}$ and $\bar{D^{0}}$ mesons required in Eq.(\ref{Eq_real_opt1}) to study the $D$ mesic nuclei are obtained
in the present work by extending the chiral SU(3) model to the SU(4) case \cite{Mishra:2008cd,Tolos:2005ft}.
The mesons $D^{0}$ and $\bar{D^{0}}$
are the members of $D$ $\left(D^{+}, D^{0}\right)$ and $\bar{D}$ 
$\left(D^{-}, \bar{D^{0}}\right)$ doublets, respectively. From the  Lagrangian density describing the interactions of $D$ and $\bar{D}$ mesons
with nucleons in the nuclear medium, equations of motion are obtained whose Fourier transformation will give the dispersion relation
	\begin{equation}
		-\omega^2+ { \textbf{k}}^2 + m_{D^{0},\bar{D^{0}}}^{2} + \Pi_{D^{0},\bar{D^{0}}}(\omega, | \textbf{k}|, \rho_i(r))=0.
		\label{drk}
	\end{equation}
For the $D^{0}$ mesons, the expression of self-energy is given by	
\begin{align} 
\Pi_{D^0}(\omega,|\textbf{k}|, \rho(r)) &  =\frac{1}{2 f_D^2}\left(2\rho_p+\rho_n\right)   \omega  +\frac{m_D^2}{2 f_D}\left(\sigma^{\prime}+\sqrt{2} \zeta_c^{\prime} + \delta^{\prime}\right) \nonumber\\ & +\left[-\frac{1}{f_D}\left(\sigma^{\prime}+\sqrt{2} \zeta_c^{\prime} + \delta^{\prime}\right)+\frac{d_1}{2 f_D^2}\left(\rho_p^s+\rho_n^s\right)\right.
  \left.+\frac{d_2}{2 f_D^2} \rho_p^s\right]\left(\omega^2-\textbf{k}^2\right).
  \label{Eq_D0_self}
\end{align}

For $\bar{D^0}$, we have
\begin{align}
 \Pi_{\bar{D^0}}(\omega,|\textbf{k}|, \rho(r)) & =-\frac{1}{2 f_D^2} \left(2\rho_p+\rho_n\right)    \omega   +\frac{m_D^2}{2 f_D}\left(\sigma^{\prime}+\sqrt{2} \zeta_c^{\prime} + \delta^{\prime}\right)\nonumber \\ & +\left[-\frac{1}{f_D}\left(\sigma^{\prime}+\sqrt{2} \zeta_c^{\prime} + \delta^{\prime}\right)+\frac{d_1}{2 f_D^2}\left(\rho_p^s+\rho_n^s\right)   +\frac{d_2}{2 f_D^2}\rho_p^s   
 \right]\left(\omega^2-\textbf{k}^2\right).
 \label{Eq_D0bar_self}
\end{align}
In Eqs. (\ref{Eq_D0_self}) and (\ref{Eq_D0bar_self}),  $\zeta_c^{\prime} $
denote the fluctuations of charm condensate $( c\bar{c})$ from the vacuum expectation
value and being heavy in flavor,  is considered zero in the current work. 
From the dispersion relation, the effective mass for the $D$ mesons (at zero momentum) is written as
\begin{equation}
m_{D^{0}, \bar{D^0}}^{*} = \sqrt{m_{D^{0}, \bar{D^0}}^{2} + 		\Pi_{D^{0}, \bar{D^0}} (\omega, | \textbf{k}|, \rho_i(r))} 
\label{Eq_mass_shiftD0}
\end{equation}
The in-medium mass of $D$ mesons  calculated using the above relation is used in Eqs. (\ref{Eq_real_opt1}) and (\ref{Eq_imga_decay1}), to obtain the real and imaginary optical potentials.

\subsection{Pseudoscalar $B^{0}$ and $\bar{B^{0}}$ mesons}
To study the formation of bound states corresponding to $B^{0}$ and $\bar{B^{0}}$ mesons chiral SU(3) model is generalized to SU(5) sector as these bottom mesons are composed of one heavy $b$  quark/antiquark.
The pseudoscalar $B^0$ and $\bar{B^{0}}$ mesons belong to the open bottom $B(B^+,B^0)$ and $\bar{B} (B^-, \bar{B^0})$ meson doublets.
In Ref. \cite{Pathak:2014nfa} the masses of open bottom mesons have been calculated in nuclear and strange hadronic matter using chiral effective model.
For our present calculations, we need the mass shift of $B^{0}$ and $\bar{B^0}$ mesons in the nuclear medium at zero temperature obtained by solving the dispersion relation
\begin{equation}
-\omega^2+ { \textbf{k}}^2 + m_{B^{0},\bar{B^{0}}}^{2} + \Pi_{B^{0},\bar{B^{0}}}(\omega, | \textbf{k}|, \rho_i(r))=0.
		\label{brk}
	\end{equation}
In case of $B^{0}$ and $\bar{B^{0}}$ mesons, the
expressions of self-energies are  given by	
 \cite{Pathak:2014nfa}
\begin{align} 
\Pi_{B^0}(\omega,|\textbf{k}|, \rho(r)) &  =-\frac{1}{2 f_B^2}\left(\rho_p+2\rho_n\right)   \omega  +\frac{m_B^2}{2 f_B}\left(\sigma^{\prime}+\sqrt{2} \zeta_b^{\prime} - \delta^{\prime}\right) \nonumber\\ & +\left[-\frac{1}{f_B}\left(\sigma^{\prime}+\sqrt{2} \zeta_b^{\prime} - \delta^{\prime}\right)+\frac{d_1}{2 f_B^2}\left(\rho_p^s+\rho_n^s\right)\right.
  \left.+\frac{d_2}{2 f_B^2} \left(\rho_p^s + 2\rho_n^s\right)\right]\left(\omega^2-\textbf{k}^2\right),
  \label{Eq_B0_self}
\end{align}
and
%For $\bar{B^0}$ meson, we have
% \cite{Pathak:2014nfa}
\begin{align} 
\Pi_{\bar{B^0}}(\omega,|\textbf{k}|, \rho(r)) &  =\frac{1}{2 f_B^2}\left(\rho_p+2\rho_n\right)   \omega  +\frac{m_B^2}{2 f_B}\left(\sigma^{\prime}+\sqrt{2} \zeta_b^{\prime} - \delta^{\prime}\right) \nonumber\\ & +\left[-\frac{1}{f_B}\left(\sigma^{\prime}+\sqrt{2} \zeta_b^{\prime} - \delta^{\prime}\right)+\frac{d_1}{2 f_B^2}\left(\rho_p^s+\rho_n^s\right)\right.
  \left.+\frac{d_2}{2 f_B^2} \left(\rho_p^s + 2\rho_n^s\right)\right]\left(\omega^2-\textbf{k}^2\right),
  \label{Eq_B0bar_self}
\end{align}
respectively.
Similar to $\zeta_c^{\prime}$, the fluctuations $\zeta_b^{\prime}$ (corresponding the condensate, $b\bar{b}$, of heavy $b$ quark)  appearing 
in Eqs. (\ref{Eq_B0_self}) and (\ref{Eq_B0bar_self}),  
are considered zero in the calculations of mass shift of  $B^0$ and $\bar{B^{0}}$ mesons. 
At zero momentum, the in-medium masses of neutral $B$ mesons required to obtain the mass shift and hence, the optical potentials in different nuclei, can be written as
\begin{equation}
m_{B^{0}, \bar{B^0}}^{*} = \sqrt{m_{B^{0}, \bar{B^0}}^{2} + 		\Pi_{B^{0}, \bar{B^0}} (\omega, | \textbf{k}|, \rho_i(r))}.
\label{Eq_mass_shiftB0}
\end{equation}
%The above calculated effective mass  of $B^0$  and $\bar{B^{0}}$ mesons will be utilized in Eqs. (\ref{Eq_real_opt1}) and (\ref{Eq_imga_decay1}), to obtain the real and imaginary optical potentials.

\subsection{Pseudoscalar  $\bar{K^{0}}$ mesons}
The in-medium masses of kaons and antikaons have been studied using chiral SU(3) model in nuclear \cite{Mishra:2006wy,Mishra:2008kg} and strange hadronic medium \cite{Mishra:2008dj}. In the nuclear matter, the effective mass of kaons $(K^+, K^0)$ is observed to increase whereas the mass of antikaons $(K^-, \bar{K^0})$ decreases as a function of density of the nuclear medium.
Since the negative mass shift implies the attractive optical potentials are required to form the bound states with nuclei, here we study the  possible formation of bound states of $\bar{K^0}$ mesons with light and heavy nuclei.
Similar to Eqs. (\ref{drk}) and (\ref{brk}), within the chiral model, the effective mass of $\bar{K^0}$ mesons will be obtained by solving the dispersion relation, for which the required self energy $\Pi_{\bar{K^0}}(\omega,|\textbf{k}|, \rho(r))$ is written as 
\cite{Mishra:2008kg}
\begin{align} 
\Pi_{\bar{K^0}}(\omega,|\textbf{k}|, \rho(r)) &  =\frac{1}{2 f_K^2}\left(\rho_p+2\rho_n\right)   \omega  +\frac{m_K^2}{2 f_K}\left(\sigma^{\prime}+\sqrt{2} \zeta^{\prime} - \delta^{\prime}\right) \nonumber\\ & +\left[-\frac{1}{f_K}\left(\sigma^{\prime}+\sqrt{2} \zeta^{\prime} - \delta^{\prime}\right)+\frac{d_1}{2 f_K^2}\left(\rho_p^s+\rho_n^s\right)\right.
  \left.+\frac{d_2}{2 f_K^2}  \rho_n^s\right]\left(\omega^2-\textbf{k}^2\right).
  \label{Eq_K0bar_self}
\end{align}

\subsection{Vector $\phi$ mesons}
\label{Subsec_phi}

To investigate the possibility of the formation of $\phi$ mesic nuclei, we calculate the effective masses and decay width of $\phi$ mesons using the effective Lagrangian  which takes into account the interactions of $\phi$ mesons with kaons, $K\left(K^{+},K^{0}\right)$ and antikaons, $\bar{K}\left(K^{-}, \bar{K^{0}}\right)$ isospin doublets. This interaction Lagrangian density is written as
\cite{Ko:1992tp,Cobos-Martinez:2017vtr} 
\begin{equation}
	\mathcal{L}_{\text {int }}= 
	i g_\phi \phi^\mu\left[\bar{K}\left(\partial_\mu K\right)-\left(\partial_\mu \bar{K}\right) K\right]. 
	\label{Eq_int_phiK1}
\end{equation}
Using above, $\phi$ meson self-energy, $\Pi_{\phi}^{*}(p)$, 
in the nuclear medium relevant for different nuclei, is calculated for the decay process $\phi \rightarrow K \bar{K}$ at the one-loop 
%(as shown in Figure \ref{fig:1}).
In Eq. (\ref{Eq_int_phiK1}), $g_{\phi}$ represents the coupling constant. 
%\begin{figure}[h]
%	\centering
%	\includegraphics[width=0.75\textwidth]{intr.jpeg}
%	\caption{The one loop level interaction of $\phi K \bar{K}$ \cite{Cobos-Martinez:2017vtr}.\label{fig:1}}
%\end{figure}
Since the contributions of $\phi \phi$K$\Bar{K}$ interactions to the in-medium masses and decay width are smaller than those of $\phi$K$\Bar{K}$ interactions, we have not taken into account these interactions in our current work on the interaction Lagrangian \cite{Cobos-Martinez:2017vtr}.
%
% The scalar component of the self-energy from the loop diagram (\ref{fig:1}) in the rest frame of the $\phi$ meson can be expressed as  
%\begin{equation}
%	i \Pi_{\phi}^{*}(p)=-\frac{8}{3} g_{\phi}^{2} \int \frac{d^{4} q}{(2 \pi)^{4}} \vec{q }^{2} D_{K}(q) D_{\bar{K}}(q-p), \label{32}
%\end{equation}
%where, $D_{K}(q)=\left(-m_{K}^{*^{2}} + q^{2} +i \epsilon\right)^{-1}$ and $D_{\bar{K}}(q-p)=\left(-m_{\bar{K}}^{*^{2}} + (q-p)^{2} +i \epsilon\right)^{-1}$ represent the propagators for kaons and antikaons, respectively. Here, $p=\left(p^{0}=m_{\phi}^{*}, \overrightarrow{0}\right)$ denotes the four-momentum vector of the $\phi$ meson, where $m_{\phi}^{*}$ indicates the in-medium mass of the $\phi$ meson. The masses of the kaon and antikaon are denoted by $m_{K}^{*}\left(=\frac{m_{K^{0}}^{*} + m_{K^{+}}^{*}}{2}\right)$ and $m_{\bar{K}}^{*}\left(=\frac{m_{\Bar{K}^{0}}^{*}+m_{K^{-}}^{*}}{2}\right)$, respectively. The values of $m_{K^{+}}^{*}, m_{K^{0}}^{*}, m_{K^{-}}^{*}$, and $m_{\bar{K}^{0}}^{*}$ are computed using Eq. (\ref{Eq_dispertion}) in finite volume strange hadronic medium are used for the calculations. 
The in-medium mass of the $\phi$ meson is expressed in terms of  
 real part of self energy, $\Pi_{\phi}^{*}(p)$,  through the relation
\begin{equation}
	m_{\phi}^{*^{2}}=\left(m_{\phi}^{0}\right)^{2}+\operatorname{Re} \Pi_{\phi}^{*}\left(m_{\phi}^{*^{2}}\right).
	 \label{Eq_phi_real_self}
\end{equation}
In the above equation, $m_{\phi}^{0}$ is the bare mass of the $\phi$ meson. Also, the real part of the self energy 
is expressed in terms of in-medium energies of kaons and antikaons as
 \cite{Kumar:2020vys,Cobos-Martinez:2017vtr}  
\begin{equation}
	\operatorname{Re} \Pi_{\phi}^{*}=-\frac{4}{3} g_{\phi}^{2} \mathcal{P} \int \frac{d^{3} q}{(2 \pi)^{3}} \vec{q}^{2} \frac{\left(E_{K}^{*}+E_{\bar{K}}^{*} \right)}{E_{K}^{*} E_{\bar{K}}^{*}\left(\left(E_{K}^{*}+E_{\bar{K}}^{*}\right)^{2}-m_{\phi}^{*^{2}}\right)}, 
	\label{Eq_real_self2_phiK}
\end{equation}
where $\mathcal{P}$ denotes the principal value of the integral. Also, $E_{K}^{*}=$ $\left(\vec{q}^{\,2}+m_{K}^{*^{2}}\right)^{1 / 2}$, and $E_{\bar{K}}^{*}=\left(\vec{q}^{\,2}+m_{\bar{K}}^{*^{2}}\right)^{1 / 2}$, where,
$m_{K}^{*}\left(=\frac{m_{K^{0}}^{*} + m_{K^{+}}^{*}}{2}\right)$ and $m_{\bar{K}}^{*}\left(=\frac{m_{\Bar{K}^{0}}^{*}+m_{K^{-}}^{*}}{2}\right)$
are the in-medium average masses of kaon and antikaon doublet, respectively.
  The values of effective masses, $m_{K^{+}}^{*}, m_{K^{0}}^{*}, m_{K^{-}}^{*}$, and $m_{\bar{K}^{0}}^{*}$ are calculated in the nuclear medium using the chiral SU(3) model \cite{Mishra:2008kg,Mishra:2008dj}.
The integral in Eq. (\ref{Eq_real_self2_phiK}) is regularized by incorporating a phenomenological form factor with the cutoff parameter $\Lambda_{c}$, i.e., \cite{Krein:2010vp} 

\begin{align}
	\operatorname{Re} \Pi_{\phi}^{*}=  -\frac{4}{3} g_{\phi}^{2} \mathcal{P} \int_{0}^{\Lambda_{c}} \frac{d^{3} q}{(2 \pi)^{3}} \vec{q}^{2}\left(\frac{\Lambda_{c}^{2}+m_{\phi}^{*^{2}}}{\Lambda_{c}^{2}+4 E_{K}^{*^{2}}}\right)^{4} 
	\frac{\left(E_{K}^{*}+E_{\bar{K}}^{*}\right)}{E_{K}^{*} E_{\bar{K}}^{*}\left(\left(E_{K}^{*}+E_{\bar{K}}^{*}\right)^{2}-m_{\phi}^{*^{2}}\right)}. \label{real_part_of_self_eng_phi}
\end{align} 
 For the case of $\phi$ meson,
 the value of decay width
  required in Eq. (\ref{Eq_imag_decay_W1}), is obtained from the imaginary component of its self energy. The expression of decay width, expressed in terms of in-medium masses of $\phi$ mesons as well as the masses of kaons and antikaons, is given by \cite{Li:1994cj}
\begin{align}
	\Gamma_{\phi}^{*}=  \frac{g_{\phi}^{2}}{24 \pi} \frac{1}{m_{\phi}^{*^{5}}}\left[\left(m_{\phi}^{*^{2}}-\left(m_{K}^{*}+m_{\bar{K}}^{*}\right)^{2}\right)\right.
	\left.\times\left(m_{\phi}^{*^{2}}-\left(m_{K}^{*}-m_{\bar{K}}^{*}\right)^{2}\right)\right]^{3 / 2}. \label{phi_decay_width}
\end{align}
The value of coupling constant $g_{\phi}$ is fixed to empirical width of the $\phi$ meson in the vacuum and found to be $4.539$.
Also,  the bare mass of the $\phi$ meson appearing in Eq. (\ref{Eq_phi_real_self}) is calculated by setting  its vacuum mass at 1019.461 MeV.

\section{Results and discussion}  \label{sec:results}
We now discuss the possibility of the formation of mesic nuclei of various mesons through the numerical calculations of the optical potentials and binding energies.
We shall calculate the optical potentials of 
$\eta, \eta^{'}$, $D^{0}$,
$\bar{D^{0}}$, $B^0$, $\bar{B^0}$,$\bar{K^{0}}$ and $\phi$ mesons in the medium relevant for the nuclei
$^{12}$C, $^{16}$O, $^{40}$Ca and $^{208}$Pb.
As we discussed in the previous section, the in-medium properties of mesons are calculated using the chiral SU(3) hadronic mean field model and extended to SU(4) and SU(5) sector to obtain the in-medium masses of charmed and bottom mesons \cite{Papazoglou:1998vr}.
The parameters $k_0$, $k_2$ and $k_4$ of chiral SU(3) model appearing in Eq. \ref{Eq_meson_self_scale1} are fitted 
to reproduce the
 vacuum values of $\sigma, \zeta$ and  $\chi$ fields \cite{Papazoglou:1998vr}.
 Further, the 
  vacuum values $\sigma_0$ and $\zeta_0$
  are calculated using the values of  
  pion decay constant $f_\pi = 93.3$ and kaon decay constant $f_K = 122$ MeV through relations $\sigma_0 = -f_\pi$ and $\zeta_0 = -\frac{1}{\sqrt{2}}(2f_K -f_\pi)$.
  The vacuum value of $\chi_0$ is constrained so that the binding energy $-16$ MeV  for symmetric nuclear matter at saturation density $\rho_0 =0.15$ fm$^{-3}$ 
  is reproduced.
  Also, the parameter $k_3$ is constrained  through the mass of $\eta$ and $\eta^{'}$ meson
 and $k_1$ is fixed to give a mass of the order of $m_\sigma \sim$
 $500$ MeV.
 The values of parameters, $d_1 = 2.56/m_K$ and
  $d_2 = 0.73/m_K$  are fitted
  to empirical values of kaon-nucleon scattering length \cite{Barnes:1992ca,Mishra:2006wy,Mishra:2008kg}.

%{\color{red}The parameters of the chiral SU(3) model used in the present calculations are given in Table \ref{table_para1}.}

\begin{table}
\begin{tabular} {|c|c|c|c|c|c|c|c|}
\hline
\multirow{2}{*}{$A$} & $nl$  & \multicolumn{3}{|c|}{ $\cal E_B $ (MeV)} & \multicolumn{3}{|c|}{$\Gamma$ (MeV)} \\
\cline{3-8}
 & & $\kappa$ =0 & $\kappa$ = 0.5 & $\kappa$ =1 & $\kappa$ =0 & $\kappa$ =  0.5 & $\kappa$ =1 \\
\hline
\multirow{1}{*}{$^{12}_{6}$C} & $1s$  & -0.511  & -0.103  & -- & 0.0 & 2.142 & --  \\
\hline
\multirow{1}{*}{$^{16}_{8}$O} & $1s$  & -1.72  & -1.38  & -0.458 & 0.0 & 3.775 & 8.094  \\
\hline
\multirow{1}{*}{$^{40}_{20}$Ca} & $1s$  & -9.769  & -9.538  & -8.91 & 0.0 & 9.527 & 19.387  \\
\hline
\multirow{6}{*}{$^{208}_{82}$Pb} & $1s$  & -20.642  & -20.555  &- 20.309 & 0.0 & 12.987 & 26.046  \\
\cline{2-8}
& $1p$  & -14.444  & -14.303  & -13.921 & 0.0 & 11.996 & 24.175  \\
\cline{2-8}
&  $1d$  & -7.244  & -7.005  & -6.392 & 0.0 & 10.621 & 21.656  \\
\cline{2-8}
&  $2s$  & -5.123  & -4.74  & -3.846 & 0.0 & 9.544 & 19.896  \\
\hline
\end{tabular}
\caption{Values of binding energies, $\cal E_B $, and full decay width, $\Gamma$, for $\eta$ mesons in four nuclei with different mass number $A$ are tabulated for $\kappa = 0, 0.5$ and 1.}
\label{table_eta_BE1}
\end{table}

\begin{figure}
\centering
\includegraphics[width=15cm, height=16cm]{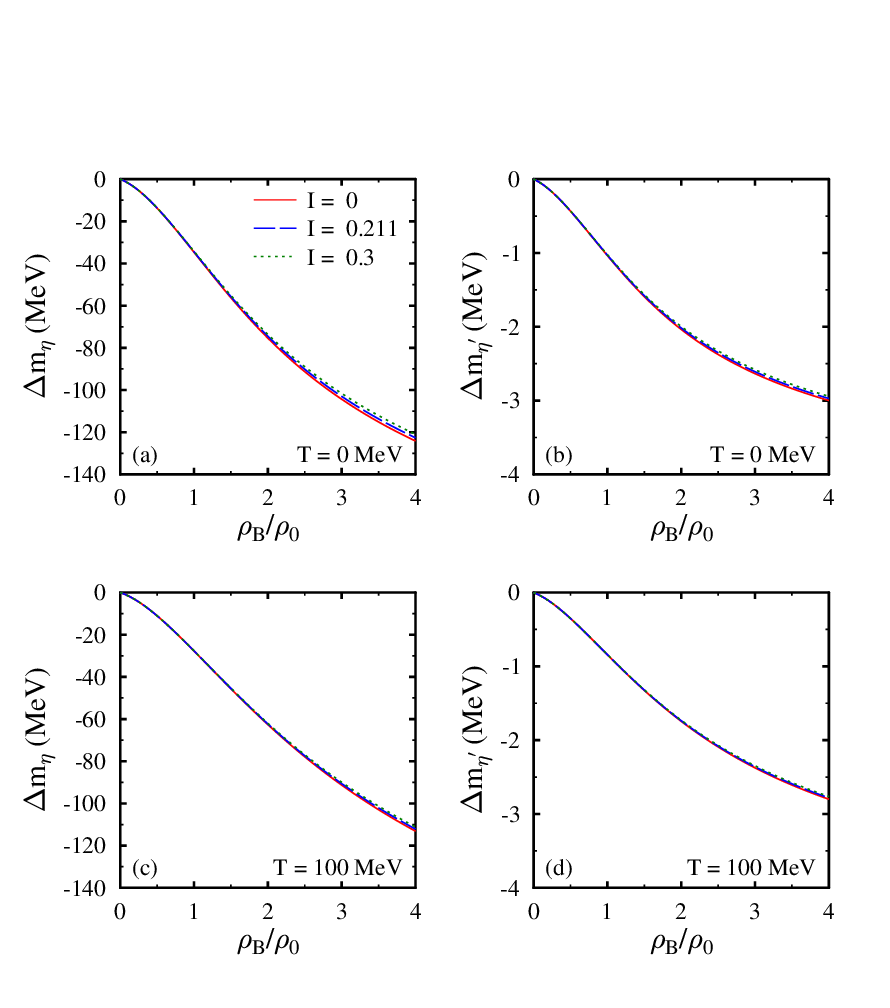}
%\vspace{-2cm}
\caption{ The mass shift of $\eta$ [in subplots (a) and (c)] and $\eta^{\prime}$ [in subplots (b) and (d)] mesons are shown as a function of density $\rho_B$ (in units of nuclear saturation density $\rho_0$) of nuclear medium for isospin asymmetry, $I = 0, 0.211$ and $0.3$. Results are shown for temperatures $T = 0 $ and $100$ MeV. \label{fig:etamass}}
\end{figure}

\begin{figure}
\centering
\includegraphics[width=15cm, height=12cm]{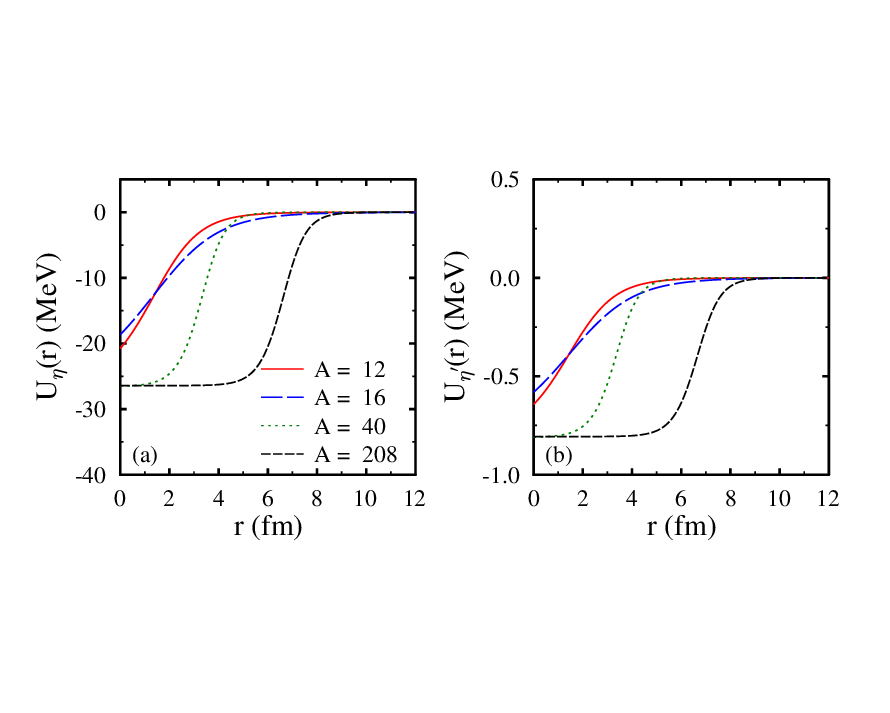}
\vspace{-2cm}
\caption{ The real optical potential $U(r)$  of $\eta$ [in subplot (a)] and $\eta^{\prime}$ [in subplot (b)] mesons are shown as a function of distance $r$ from center, for nuclei
$^{12}$C, $^{16}$O, $^{40}$Ca and $^{208}$Pb, with mass number, $A = 12, 16, 40$ and $208$, respectively. \label{fig:eta_optical}}
\end{figure}

\subsection{Binding energies for $\eta$ and $\eta^{'}$ mesic nuclei}
 In this subsection, we shall explore the possibility of the formation of $\eta$ and $\eta^{'}$ mesic nuclei.
 Using Eq. (\ref{Eq_mass_matrix_eta}),
 the mass shifts of pseudoscalar $\eta$ and $\eta^{'}$ mesons are calculated within the chiral SU(3) model. 
 In Fig. \ref{fig:etamass},
 we have shown the mass shift of $\eta$ and $\eta^{'}$ mesons as a function of baryon density $\rho_B$ (in units of nuclear saturation density $\rho_0$) of the infinite nuclear medium for isospin 
 asymmetry $I = 0, 0.211 $ and $0.3$, at temperatures $T = 0$ and $100$ MeV. The isospin asymmetry $ I = 0.211$ corresponds to the heavy nuclei $^{208}$Pb considered in the present work for the bound state formation.
 As we discussed earlier, for the calculations of $\eta$ and $\eta^{'}$ meson
 properties, we considered the mixing of $\eta_8$ and $\eta_0$ states.
  The pseudoscalar $\eta$ meson which is member of octet is observed to undergo larger mass shift as compared to the singlet $\eta^{'}$.
 At nuclear saturation density $\rho_B = \rho_0$,
for isospin asymmetry $I = 0(0.3)$ and temperature $T = 0$ MeV, the values of mass shift 
 $\Delta m_\eta$
 and $\Delta m_\eta^{'}$ are observed to be $-34.64 (-34.21)$ and $-1.03 (-1.02)$ MeV, respectively. 
 At baryon density $\rho_B= 4\rho_0$, the above values of mass shift change  to
 $-124.15(-120.92)$ and $-3.00 (-2.94)$ MeV.
 The finite isopsin asymmetry of the medium causes less drop in the masses of $\eta$ and $\eta^{'}$, although the impact is small.
 An increase in the temperature of nuclear medium from $T = 0$ to $100$ MeV, enhances the
 in-medium mass of these mesons.
 Furthermore, in the present chiral model calculations, compared to $\eta$ mesons, the singlet $\eta^{'}$ undergoes very less drop in mass as a function of density of nuclear medium. Recall that the in-medium masses of $\eta$ and $\eta^{'}$ mesons are
 obtained from the diagonalization of Eq. (\ref{Eq_mass_matrix_eta}). Neglecting the off-diagonal terms (having very small contributions),  $m_{\eta}^{*} \sim m_{\eta_8}^{*}$  and $m_{\eta{'}}^{*} \sim m_{\eta_0}^{*}$. As can be seen from Eq.(\ref{Eq_meta0_med1}), within the chiral SU(3) model $m_{\eta_0}^{*} = m_{\eta_0}$, i.e., $\eta_0$  is not modified in the medium and therefore, $\eta^{'}$ mesons
 are not modified much in the present calculations (small modifications are stimulated through the diagonalization of Eq. (\ref{Eq_mass_matrix_eta}).
 
 To study the bound states, the Klein Gordon equation
 [Eq. (\ref{Eq_KG_Coord1})], which require the optical potential corresponding to different nuclei under study as input is solved. 
 In Fig. 
 \ref{fig:eta_optical}
 the real part of the optical potentials, for $\eta$ and $\eta^\prime$ mesons,
 calculated using 
 Eq. (\ref{Eq_real_opt1})
 is plotted as a function of distance $r$ from the center of nuclei.
 The heavy nuclei are observed to have larger value of negative optical potential values as compared
 to lighter nuclei.
The imaginary optical potential which corresponds to the absorption of mesons in the nuclei is calculated using Eq. (\ref{Eq_imag_decay_W1}).
Using the real and imaginary values of optical potential in the Klein Gordon equation, the binding energy and absorption decay width are calculated.
In case of $\eta$ mesons,  the values of binding energies and decay width are tabulated in Table \ref{table_eta_BE1}. 
In case of light nuclei $^{12}$C the negative values for the binding energies are observed only in $1s$ state
for the value of phenomenological parameter $\kappa = 0$ and $0.5$.
As described earlier 
in Sec. \ref{sec:potential}, through the parameter $\kappa$ we simulate the impact of imaginary optical potential on the formation of bound states.
As the value of 
$\kappa$ is increased to $1$, no bound state is observed.
For the case of $^{16}$O and $^{40}$Ca, bound state formation in the $1s$ state may take place for all values of $\kappa$ considered in the present work.
As can be seen from Table \ref{table_eta_BE1}, in the case of $^{208}$Pb nuclei the negative values for the binding energies are observed not only for the $1s$ state, but also for the other excited states, for example, $1p, 1d$ and $2s$ as well. Due to larger value of attractive optical potential (magnitude) of $\eta$ meson with heavy $^{208}$Pb nuclei, the possibility of the formation of bound state is more with these nuclei.

As discussed earlier, in the present calculations the $\eta^{'}$ mesons are not observed to undergo appreciable  mass drop in the nuclear medium ($\Delta m_{\eta^{'}} = -1.03$ MeV at $\rho_B = \rho_0$) and
bound states are not observed for these pseudoscalar singlet meson with any of the nuclei considered in the present work. The present results of $\eta$ and $\eta^{'}$ mesons can be compared with the earlier studies using different approaches.
In Ref. \cite{Cobos-Martinez:2023hbp} the mass shifts of $\eta$ and $\eta^{'}$ mesons were calculated in the symmetric nuclear medium using the QMC model considering the $\eta-\eta^{'}$ mixing.
The values of mass shift are observed to be $-63.6$  and $-61.3$ MeV, respectively, at $\rho_B =\rho_0$.
The binding energies and absorption decay width were calculated  for light and heavy nuclei and were observed to form the bound states due to significant mass drop.
Using the coupled channel approach, the optical potential for $\eta^{'}$ mesons was found to be
$-(8.7 + 1.8 i)$ MeV at nuclear saturation density, $\rho_0$ and no bound state is observed in Ref. \cite{Nagahiro:2011fi}.
 Bound states of $\eta$ mesons  have also been explored using unitarized coupled channel approach in 
 Ref. \cite{Garcia-Recio:2002xog}, considering the energy independent and dependent potentials.
 The binding energies obtained were larger compared to our present calculations.
 The feasibility of the formation of $\eta$ and $\eta^{'}$ meson bound states with nuclei has also been explored using the NJL model \cite{Nagahiro:2006dr}.
 In Ref. \cite{Jido:2018aew}, relativistic field theoretical model under mean field approximation is used to study the $\eta^{\prime}$ mesic nuclei through the direct coupling of $\eta^{\prime}$ meson with scalar mean field $\sigma$ (interaction of form $\sim g_{\sigma \eta^{\prime}} m_{\eta^{\prime}} \eta^{\prime 2}\sigma$). Assuming the partial restoration of chiral symmetry and $30$\%
 reduction in the chiral condensate to determine the coupling constant $g_{\sigma \eta^{\prime}}$, the $\eta^{\prime}$ meson is observed to undergo a mass drop of $80$ MeV at nuclear saturation density from vacuum value.
 Such larger mass drop obviously lead to the formation of bound states with nuclei $^{12}$C, $^{16}$O and $^{40}$Ca, considered in the work.

% \begin{table}
%\begin{tabular} {|c|c|c|c|c|c|c|c|}
%\hline
%\multirow{2}{*}{A} & nl  & \multicolumn{3}{|c|}{B.E.} & \multicolumn{3}{|c|}{-$\Gamma$/2} \\
%\cline{3-8}
% & & $\gamma$ =0 & $\gamma$ = 0.5 & $\gamma$ =1 & $\gamma$ =0 & $\gamma$ =  0.5 & $\gamma$ =1 \\
%\hline
%\multirow{3}{*}{$^{12}_{6}{C}$} & 1s  & 0.069  & 0.069  & 0.069 & -0.0 & 0.0 & 0.001  \\
%\cline{2-8}
%& 1p  & 0.069  & 0.069  & 0.069 & -0.0 & 0.0 & 0.0  \\
%\cline{2-8}
%&  1d  & 0.069  & 0.069  & 0.069 & -0.0 & 0.0 & 0.0  \\
%\hline
%\multirow{3}{*}{$^{16}_{8}{O}$} & 1s  & 0.067  & 0.067  & 0.067 & -0.0 & 0.0 & 0.001  \\
%\cline{2-8}
%& 1p  & 0.068  & 0.068  & 0.068 & -0.0 & 0.0 & 0.0  \\
%\cline{2-8}
%&  1d  & 0.068  & 0.068  & 0.068 & -0.0 & 0.0 & 0.0  \\
%\hline
%\multirow{3}{*}{$^{40}_{20}{Ca}$} & 1s  & 0.063  & 0.063  & 0.063 & -0.0 & 0.001 & 0.003  \\
%\cline{2-8}
%& 1p  & 0.065  & 0.065  & 0.065 & -0.0 & 0.0 & 0.0  \\
%\cline{2-8}
%&  1d  & 0.065  & 0.065  & 0.065 & -0.0 & 0.0 & 0.0  \\
%\hline
%\multirow{3}{*}{$^{208}_{82}{Pb}$} & 1s  & 0.04  & 0.043  & 0.051 & -0.0 & 0.023 & 0.04  \\
%\cline{2-8}
%& 1p  & 0.064  & 0.064  & 0.064 & -0.0 & 0.0 & 0.0  \\
%\cline{2-8}
%&  1d  & 0.064  & 0.064  & 0.064 & -0.0 & 0.0 & 0.0  \\
%\hline
%\end{tabular}
%\caption{$\eta^{'}$}
% \end{table}

\begin{table}
\begin{tabular} {|c|c|c|c|c|c|c|c|}
\hline
\multirow{2}{*}{$A$} & $nl$  & \multicolumn{3}{|c|}{$\cal E_B$ (MeV)} & \multicolumn{3}{|c|}{$\Gamma$ (MeV)} \\
\cline{3-8}
 & & $\kappa$ =0 & $\kappa$ = 0.5 & $\kappa$ =1 & $\kappa$ =0 & $\kappa$ =  0.5 & $\kappa$ =1 \\
\hline
\multirow{3}{*}{$^{12}_{6}$C} & 1s  & -25.965  & -25.741  & -25.118 & 0.0 & 18.731 & 37.748  \\
\cline{2-8}
& 1p  & -11.246  & -10.811  & -9.638 & 0.0 & 14.235 & 29.144  \\
\cline{2-8}
&  2s  & -0.935  & --  & -- & 0.0 & -- & --  \\
\hline
\multirow{4}{*}{$^{16}_{8}$O} & 1s  & -29.11  & -28.941  & -28.474 & 0.0 & 19.504 & 39.224  \\
\cline{2-8}
& 1p  & -16.516  & -16.184  & -15.282 & 0.0 & 16.451 & 33.407  \\
\cline{2-8}
&  1d  & -3.884  & -3.234  & -1.619 & 0.0 & 12.057 & 25.31  \\
\cline{2-8}
&  2s  & -3.946  & -3.152  & -1.182 & 0.0 & 9.717 & 20.971  \\
\hline
\multirow{6}{*}{$^{40}_{20}$Ca} & 1s  & -47.743  & -47.607  & -47.22 & 0.0 & 27.865 & 55.842  \\
\cline{2-8}
& 1p  & -38.223  & -38.016  & -37.44 & 0.0 & 26.065 & 52.369  \\
\cline{2-8}
&  1d  & -27.642  & -27.338  & -26.509 & 0.0 & 23.945 & 48.313  \\
\cline{2-8}
&  2s  & -25.594  & -25.233  & -24.263 & 0.0 & 23.249 & 47.047  \\
\cline{2-8}
&  2p  & -14.212  & -13.608  & -12.07 & 0.0 & 19.93 & 40.966  \\
\cline{2-8}
&  2d  & -3.852  & -2.7  & -0.152 & 0.0 & 15.302 & 33.133  \\
\hline
\multirow{6}{*}{$^{208}_{82}$Pb} & 1s  & -49.133  & -49.073  & -48.897 & 0.0 & 26.019 & 52.054  \\
\cline{2-8}
& 1p  & -46.417  & -46.344  & -46.135 & 0.0 & 25.734 & 51.507  \\
\cline{2-8}
&  1d  & -43.096  & -43.008  & -42.757 & 0.0 & 25.382 & 50.832  \\
\cline{2-8}
&  2s  & -41.77  & -41.674  & -41.401 & 0.0 & 25.235 & 50.554  \\
\cline{2-8}
&  2p  & -37.132  & -37.008  & -36.664 & 0.0 & 24.711 & 49.56  \\
\cline{2-8}
&  2d  & -32.078  & -31.919  & -31.486 & 0.0 & 24.102 & 48.414  \\
\hline
\end{tabular}
\caption{Values of binding energies, $\cal E_B $, and  full decay width, $\Gamma$ for $D^0$ mesons in four nuclei with different mass number $A$ are tabulated for $\kappa = 0, 0.5$ and 1.}
\label{table_D0_BE1}
\end{table}

\begin{table}
\begin{tabular} {|c|c|c|c|c|c|c|c|}
\hline
\multirow{2}{*}{$A$} & $nl$  & \multicolumn{3}{|c|}{$\cal E_B$ (MeV)} & \multicolumn{3}{|c|}{$\Gamma$ (MeV)} \\
\cline{3-8}
 & & $\kappa$ =0 & $\kappa$ = 0.5 & $\kappa$ =1 & $\kappa$ =0 & $\kappa$ =  0.5 & $\kappa$ =1 \\
\hline
\multirow{1}{*}{$^{12}_{6}$C} & 1s  & -1.633  & -1.49  & -1.102 & 0.0 & 2.539 & 5.306  \\
\hline
\multirow{1}{*}{$^{16}_{8}$O} & 1s  & -2.142  & -2.03  & -1.726 & 0.0 & 2.756 & 5.694  \\
\hline
\multirow{2}{*}{$^{40}_{20}$Ca} & 1s  & -9.87  & -9.801  & -9.615 & 0.0 & 7.096 & 14.299  \\
\cline{2-8}
& 1p  & -4.25  & -4.091  & -3.676 & 0.0 & 5.674 & 11.637  \\
\hline
\multirow{6}{*}{$^{208}_{82}$Pb} & 1s  & -13.541  & -13.524  & -13.479 & 0.0 & 7.646 & 15.313  \\
\cline{2-8}
& 1p  & -11.434  & -11.403  & -11.32 & 0.0 & 7.41 & 14.869  \\
\cline{2-8}
&  1d  & -8.887  & -8.836  & -8.701 & 0.0 & 7.101 & 14.29  \\
\cline{2-8}
&  2s  & -7.936  & -7.869  & -7.694 & 0.0 & 6.924 & 13.973  \\
\cline{2-8}
&  2p  & -4.607  & -4.478  & -4.164 & 0.0 & 6.303 & 12.881  \\
\cline{2-8}
&  2d  & -1.3  & -1.007  & -0.43 & 0.0 & 5.311 & 11.331  \\
\hline
\end{tabular}
\caption{Values of binding energies, $\cal E_B $, and full  decay width, $\Gamma$, for $\bar{D^0}$ mesons in four nuclei with different mass number $A$ are tabulated for $\kappa = 0, 0.5$ and 1.}
\label{table_D0bar_BE1}
\end{table}

\subsection{Binding energies for $D^{0}$ and $\bar{D^{0}}$ mesic nuclei}
\begin{figure}
\centering
\includegraphics[width=15cm, height=16cm]{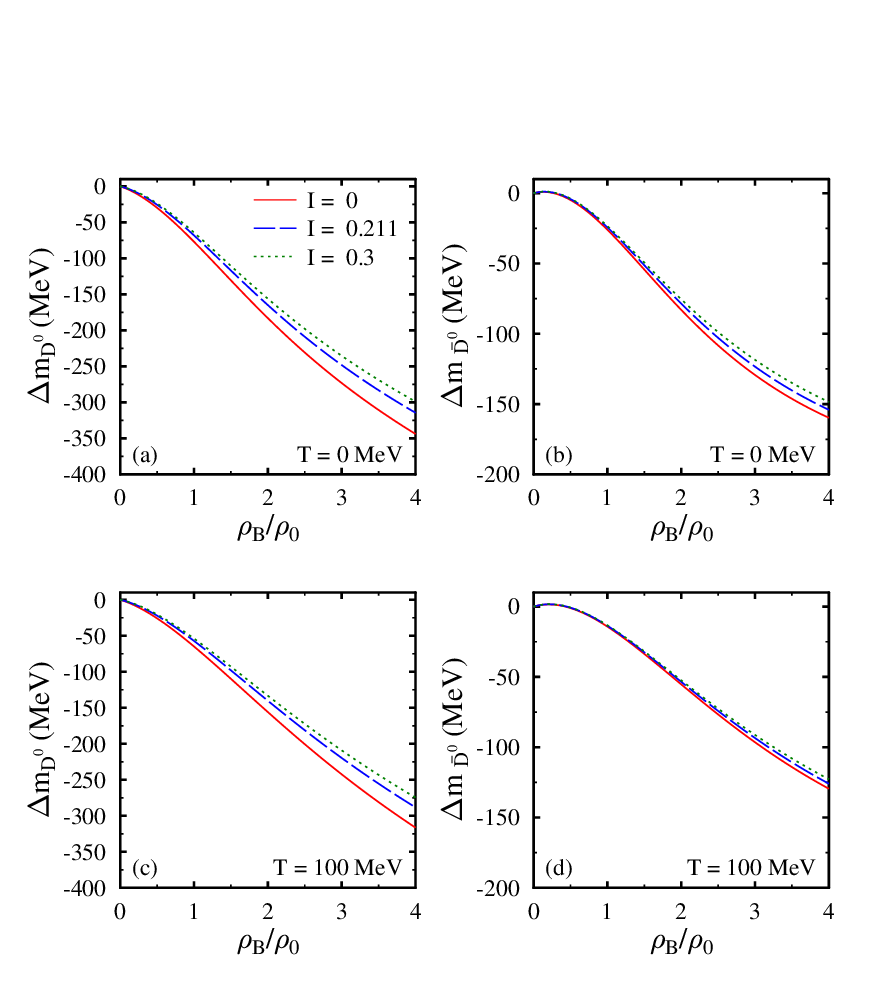}
%\vspace{-2cm}
\caption{ The mass shift of $D^{0}$ [in subplots (a) and (c)] and $\bar{D^{0}}$ [in subplots (b) and (d)] mesons are shown as a function of baryon density $\rho_B$ (in units of nuclear saturation density $\rho_0$) of nuclear medium for isospin asymmetry, $I = 0, 0.211$ and $0.3$. Results are shown for temperatures $T = 0 $ and $100$ MeV. \label{fig:Dmass}}
\end{figure}

Now we discuss the formation of bound states for neutral 
$D^{0}$ and $\bar{D^{0}}$ mesons  with the nuclei  within the chiral SU(3) hadronic mean field model.
The mass shift of $D^0$ and $\bar{D^0}$ mesons calculated using Eq. (\ref{Eq_mass_shiftD0}) are plotted in Fig. (\ref{fig:Dmass}) as a function of  baryon density ratio $\rho_B/\rho_0$.
As can be seen from the figure, the in-medium masses of both  $D^0$ and $\bar{D^0}$ mesons decrease as a function of density of the nuclear medium.
 Furthermore, the increase of isospin asymmetry of the medium from zero to finite value  (which is relevant for the study of nuclei having different number of protons and neutrons)
decreases the magnitude of mass shift for these neutral pseudoscalar mesons.
At zero temperature and  baryon density $\rho_B = \rho_0 (4\rho_0)$, the values of mass shift for $D^0$ and $\bar{D^0}$ mesons are observed to $-76.96 (-343.82)$ and $-26 (-159.71)$ MeV in the symmetric nuclear medium, whereas for $I = 0.3$ these values changes to $-64.17 (-298.80)$ and $-23.37 (-148.43)$ MeV, respectively.
For temperature $T = 100$ MeV, the values of mass shift for  $D^0$ and $\bar{D^0}$ mesons
are observed to be $-64.92(-316.43) $ and $-14.24(-129.53)$ MeV, respectively,
in symmetric nuclear medium at $\rho_B =
\rho_0 (4\rho_0)$.

\begin{figure}
\centering
\includegraphics[width=15cm, height=16cm]{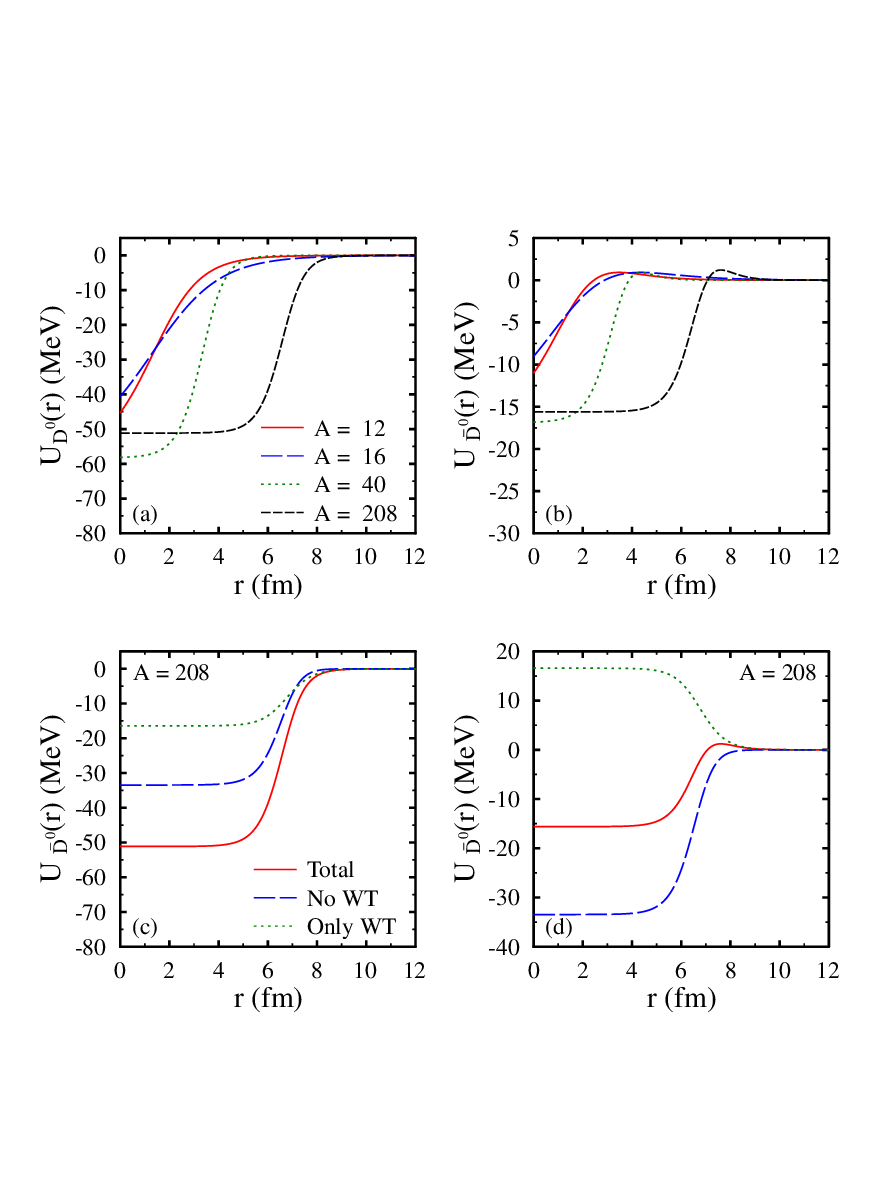}
\vspace{-2cm}
\caption{ The real optical potentials $U(r)$  of $D^0$ [in subplot (a)] and $\bar{D^{0}}$ [in subplot (b)] mesons are shown as a function of distance $r$ from center, for nuclei
$^{12}$C, $^{16}$O, $^{40}$Ca and $^{208}$Pb, with mass number, $A = 12, 16, 40$ and $208$, respectively.
In subplots (c)  and (d)  optical potentials are plotted for $A =208$ considering the impact of Weinberg Tomozawa (WT) term to the total optical potential.
 \label{fig:D_optical}}
\end{figure}

Using the observed values of mass shift for $D^{0}$ and $\bar{D^0}$ mesons at center of given nuclei in Eq. (\ref{Eq_real_opt1}),  the real part of optical potential is calculated and is plotted in   Fig. \ref{fig:D_optical} as a function of $r$ for nuclei,
$^{12}$C, $^{16}$O, $^{40}$Ca and $^{208}$Pb.
As one can see from the figure, depth of optical potential (large negative value) is observed to be more for $D^{0}$ mesons as compared to $\bar{D^{0}}$.
 In case of $\bar{D^0}$ mesons
the optical potential $U_{\bar{D^0}}(r)$ is observed to have  positive  values for certain range of  $r$ causing a bump as can be seen from Fig. \ref{fig:D_optical}(b).
On the otherside, such positive values are not observed for the optical potential of $D^{0}$ mesons. The observed difference is
because of opposite contribution  of Weinberg Tomozawa term for the
$D^{0}$ and $\bar{D^{0}}$ mesons as can be seen from Eqs. (\ref{Eq_D0_self}) and (\ref{Eq_D0bar_self}), respectively.
Weinberg Tomozawa term alone gives positive contribution to the optical potential of 
$\bar{D^{0}}$ mesons, whereas the combined contribution of all other terms give  negative value for $U_{\bar{D^0}}(r)$.
For certain range of $r$, the positive contribution due to Weinberg Tomozawa term dominates over the net negative contribution of other terms, causing overall positive value of optical potential for $\bar{D^0}$ mesons. To understand this in 
Fig. \ref{fig:D_optical}(c)  and Fig. \ref{fig:D_optical}(d), we  plot the optical potentials of $D^0$ and $\bar{D^0}$ mesons, respectively, considering (i) Weinberg Tomozawa term only  and (ii) without Weinberg Tomozawa term, for $^{208}$Pb. As can be seen from
Fig. \ref{fig:D_optical}(c), in case of $D^{0}$ mesons optical potentials from Weinberg term have negative values and therefore, unlike $\bar{D^0}$ mesons, positive values are not observed for $D^0$ mesons.

The imaginary part of optical potential, required to  understand the absorption of mesons in the nuclei and its impact on the binding energy of neutral $D$ mesic nuclei, is calculated using Eq. (\ref{Eq_imag_decay_W1}).
The values of binding energy
and decay width  obtained by solving the  
Klein Gordon equation are given in Tables \ref{table_D0_BE1} and \ref{table_D0bar_BE1}, for
$D^0$ and $\bar{D^0}$ mesons, respectively, for
$\kappa = 0, 0.5$ and 1.
As one can see from Table \ref{table_D0_BE1},  the bound states of $D^0$ mesons,
for $^{40}$Ca and $^{208}$Pb
may be found not only for lower states $1s$ and $1p$, but also for other higher states when $\kappa = 0$ and $0.5$. When the value of $\kappa$ is increased to one,
the magnitude of binding energy is observed to be less as compared to the decay width value. This may cause a broadening of the peak and hindrance in the observation of clear signature of bound states.
Comparing the values of binding energies in Tables \ref{table_D0_BE1} and \ref{table_D0bar_BE1}, 
magnitude of binding energies for $\bar{D^0}$ mesons are significantly less than $D^0$ mesons.
In Ref. \cite{Garcia-Recio:2010fiq}, the bound states of $D^{0}$ mesons have been studied for several nuclei using the coupled channel approach.
Comparing our results for
$^{12}$C, $^{40}$Ca and $^{208}$Pb with Ref. \cite{Garcia-Recio:2010fiq}, the magnitude of binding energies are observed to be more in our case.
The binding energies of $D^0$ and $\bar{D^0}$ mesons for $^{208}$Pb nuclei are calculated using QMC model and are found to approximately $ -96$ and $-25$ MeV, respectively \cite{Tsushima:1998ru}.

\begin{figure}
\centering
\includegraphics[width=15cm, height=16cm]{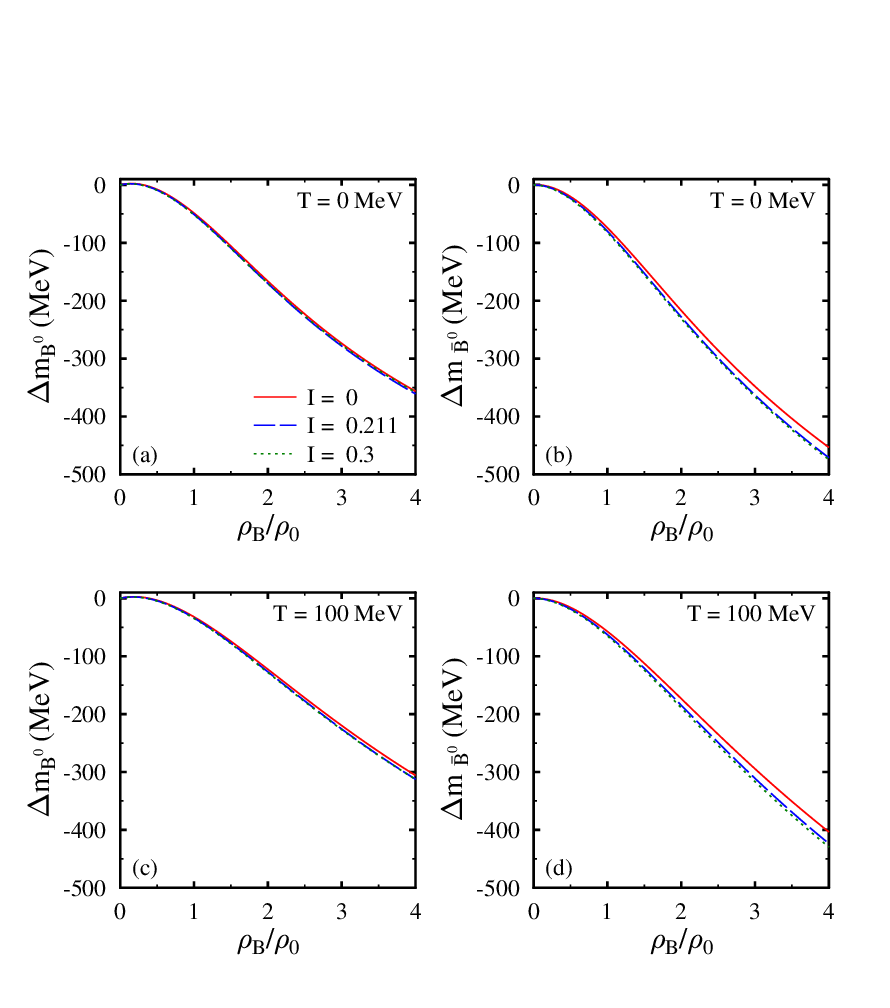}
%\vspace{-2cm}
\caption{ The mass shift of $B^{0}$ [in subplots (a) and (c)] and $\bar{B^{0}}$ [in subplots (b) and (d)] mesons are shown as a function of density $\rho_B$ (in units of nuclear saturation density $\rho_0$) of nuclear medium for isospin asymmetry, $I = 0, 0.211$ and $0.3$. Results are shown for temperatures $T = 0 $ and $100$ MeV. \label{fig:Bmass}}
\end{figure}

\begin{figure}
\centering
\includegraphics[width=15cm, height=12cm]{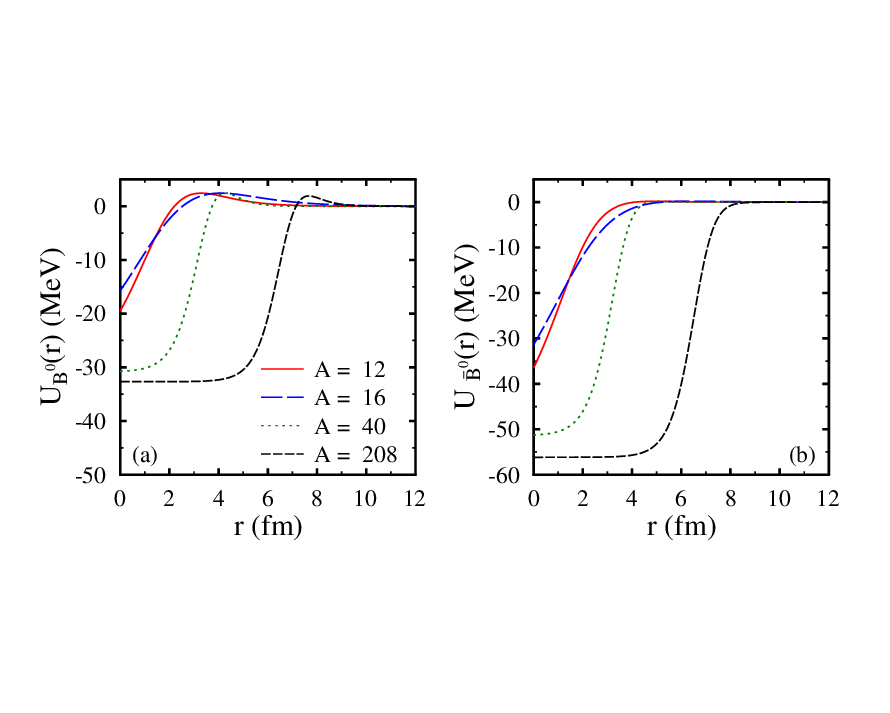}
\vspace{-2cm}
\caption{ The real optical potential $U(r)$  of $B^0$ [in subplot (a)] and $\bar{B^{0}}$ [in subplot (b)] mesons are shown as a function of distance $r$ from center, for nuclei
$^{12}$C, $^{16}$O, $^{40}$Ca and $^{208}$Pb, with mass number, $A = 12, 16, 40$ and $208$, respectively. 
 \label{fig:B_optical}}
\end{figure}
\subsection{Binding energies for $B^{0}$ and $\bar{B^{0}}$ mesic nuclei}
The studies on the  bound states of $B$ meson with nuclei has not been done much in literature \cite{Mondal:2024vyt}. 
In the present section we shall discuss the formation of bound states of pseudoscalar $B^{0}$ and $\bar{B^{0}}$ mesons by calculating their optical potential using the chiral SU(3) model.
In Fig. \ref{fig:Bmass}, the mass shift of these pseudoscalar mesons are plotted as a function of baryon density $\rho_B$ (in units of $\rho_0$) for isospin asymmetry $I = 0, 0.211$ and $0.3$, at $T = 0$ and $100$ MeV.
Both pseudoscalar $B^0$ and $\bar{B^0}$ mesons are
observed to undergo appreciable mass drop at nuclear matter density, whereas, for given density, an increase in temperature from zero to finite values causes an increase in the mass, i.e., less drop in mass is observed.
However, the isospin asymmetry effects are not appreciable. In cold symmetric nuclear medium, $\rho_B = \rho_0$, the values of $\Delta m$ for $B^0$ and $\bar{B^0}$ mesons are observed to be $-48.69$ and $-74.06$ MeV, whereas at $4\rho_0$ these values changes to $-356.40$ and $-453.28$ MeV, respectively.
As shown in Fig. \ref{fig:B_optical}, for a given nuclei, at the center, the real part of optical potentials have larger value for $\bar{B^{0}}$
as compared to $B^0$ mesons.
In case of $B^0$ mesons
the optical potentials have positive values for certain range whereas for $\bar{B^{0}}$ this is not the case. This is because of opposite contribution of Weinberg Tomozawa term to $B^0$ and $\bar{B^{0}}$ mesons
(see first term of Eqs.(\ref{Eq_B0_self}) and (\ref{Eq_B0bar_self})). For $B^0$ mesons, Weinberg Tomozawa term lead to positive values of optical potentials.
The values of binding energy and decay width for these pseudoscalar mesons are  given in Table \ref{table_B0_BE1} and \ref{table_B0bar_BE1}.
As expected from the magnitude of mass drop for $B^0$ and $\bar{B^0}$ mesons, the magnitude of binding energy with given nuclei is more for $\bar{B^0}$ mesons as compared to $B^0$.

\begin{table}
\begin{tabular} {|c|c|c|c|c|c|c|c|}
\hline
\multirow{2}{*}{$A$} & $nl$  & \multicolumn{3}{|c|}{$\cal E_B$ (MeV)} & \multicolumn{3}{|c|}{$\Gamma$ (MeV)} \\
\cline{3-8}
 & & $\kappa$ =0 & $\kappa$ = 0.5 & $\kappa$ =1 & $\kappa$ =0 & $\kappa$ =  0.5 & $\kappa$ =1 \\
\hline
\multirow{3}{*}{$^{12}_{6}$C} & 1s  & -10.933  & -10.845  & -10.604 & 0.0 & 7.907 & 15.944  \\
\cline{2-8}
& 1p  & -4.572  & -4.385  & -3.886 & 0.0 & 5.989 & 12.285  \\
\cline{2-8}
&  2s  & -0.272  & --  & -- & 0.0 & -- & --  \\
\hline
\multirow{4}{*}{$^{16}_{8}$O} & 1s  & -10.848  & -10.789  & -10.627 & 0.0 & 7.362 & 14.816  \\
\cline{2-8}
& 1p  & -5.789  & -5.657  & -5.302 & 0.0 & 6.133 & 12.486  \\
\cline{2-8}
&  1d  & -0.805  & -0.513  & -- & 0.0 & 4.274 & --  \\
\cline{2-8}
&  2s  & -1.02  & -0.662  & -- & 0.0 & 3.274 & --  \\
\hline
\multirow{6}{*}{$^{40}_{20}$Ca} & 1s  & -26.07  & -26.032  & -25.926 & 0.0 & 14.613 & 29.275  \\
\cline{2-8}
& 1p  & -22.064  & -21.996  & -21.809 & 0.0 & 13.928 & 27.954  \\
\cline{2-8}
&  1d  & -17.562  & -17.456  & -17.166 & 0.0 & 13.13 & 26.425  \\
\cline{2-8}
&  2s  & -16.627  & -16.502  & -16.166 & 0.0 & 12.902 & 26.009  \\
\cline{2-8}
&  2p  & -11.516  & -11.316  & -10.79 & 0.0 & 11.741 & 23.84  \\
\cline{2-8}
&  2d  & -6.481  & -6.158  & -5.352 & 0.0 & 10.282 & 21.205  \\
\hline
\multirow{6}{*}{$^{208}_{82}$Pb} & 1s  & -31.71  & -31.698  & -31.665 & 0.0 & 16.342 & 32.691  \\
\cline{2-8}
& 1p  & -30.648  & -30.632  & -30.586 & 0.0 & 16.245 & 32.505  \\
\cline{2-8}
&  1d  & -29.347  & -29.325  & -29.263 & 0.0 & 16.125 & 32.277  \\
\cline{2-8}
&  2s  & -28.826  & -28.802  & -28.733 & 0.0 & 16.08 & 32.19  \\
\cline{2-8}
&  2p  & -26.996  & -26.961  & -26.868 & 0.0 & 15.91 & 31.867  \\
\cline{2-8}
&  2d  & -24.983  & -24.938  & -24.815 & 0.0 & 15.717 & 31.502  \\
\hline
\end{tabular}

\caption{Values of binding energies, $\cal E_B $, and full decay width, $\Gamma$ for $B^{0}$ mesons in four nuclei with different mass number $A$ are tabulated for $\kappa = 0, 0.5$ and 1.}
\label{table_B0_BE1}
\end{table}

\begin{table}
\begin{tabular} {|c|c|c|c|c|c|c|c|}
\hline
\multirow{2}{*}{$A$} & $nl$  & \multicolumn{3}{|c|}{$\cal E_B$ (MeV)} & \multicolumn{3}{|c|}{$\Gamma$ (MeV)} \\
\cline{3-8}
 & & $\kappa$ =0 & $\kappa$ = 0.5 & $\kappa$ =1 & $\kappa$ =0 & $\kappa$ =  0.5 & $\kappa$ =1 \\
\hline
\multirow{5}{*}{$^{12}_{6}$C} & 1s  & -25.403  & -25.292  & -24.986 & 0.0 & 16.151 & 32.455  \\
\cline{2-8}
& 1p  & -16.432  & -16.222  & -15.646 & 0.0 & 13.936 & 28.186  \\
\cline{2-8}
&  1d  & -7.473  & -7.117  & -6.171 & 0.0 & 11.205 & 22.994  \\
\cline{2-8}
&  2s  & -7.506  & -7.101  & -6.028 & 0.0 & 10.381 & 21.446  \\
\cline{2-8}
&  2p  & -1.086  & -0.26  & -- & 0.0 & 6.142 & --  \\
\hline
\multirow{5}{*}{$^{16}_{8}$O} & 1s  & -26.085  & -26.014  & -25.818 & 0.0 & 15.715 & 31.532  \\
\cline{2-8}
& 1p  & -19.262  & -19.116  & -18.717 & 0.0 & 14.435 & 29.096  \\
\cline{2-8}
&  1d  & -11.721  & -11.471  & -10.799 & 0.0 & 12.632 & 25.674  \\
\cline{2-8}
&  2s  & -10.487  & -10.193  & -9.407 & 0.0 & 11.52 & 23.535  \\
\cline{2-8}
&  2p  & -4.153  & -3.645  & -2.362 & 0.0 & 8.743 & 18.447  \\
\hline
\multirow{6}{*}{$^{40}_{20}$Ca} & 1s  & -45.779  & -45.725  & -45.572 & 0.0 & 24.844 & 49.743  \\
\cline{2-8}
& 1p  & -41.003  & -40.917  & -40.677 & 0.0 & 24.034 & 48.174  \\
\cline{2-8}
&  1d  & -35.617  & -35.492  & -35.146 & 0.0 & 23.119 & 46.409  \\
\cline{2-8}
&  2s  & -34.457  & -34.317  & -33.934 & 0.0 & 22.917 & 46.034  \\
\cline{2-8}
&  2p  & -28.121  & -27.919  & -27.371 & 0.0 & 21.716 & 43.749  \\
\cline{2-8}
&  2d  & -21.612  & -21.326  & -20.57 & 0.0 & 20.34 & 41.161  \\
\hline
\multirow{6}{*}{$^{208}_{82}$Pb} & 1s  & -55.277  & -55.252  & -55.179 & 0.0 & 28.274 & 56.555  \\
\cline{2-8}
& 1p  & -54.102  & -54.073  & -53.986 & 0.0 & 28.164 & 56.343  \\
\cline{2-8}
&  1d  & -52.663  & -52.627  & -52.523 & 0.0 & 28.029 & 56.085  \\
\cline{2-8}
&  2s  & -52.09  & -52.052  & -51.941 & 0.0 & 27.981 & 55.992  \\
\cline{2-8}
&  2p  & -50.065  & -50.017  & -49.883 & 0.0 & 27.796 & 55.639  \\
\cline{2-8}
&  2d  & -47.833  & -47.775  & -47.612 & 0.0 & 27.59 & 55.244  \\
\hline
\end{tabular}

\caption{Values of binding energies, $\cal E_B $, and full decay width, $\Gamma$ for $\bar{B^0}$ mesons in four nuclei with different mass number $A$ are tabulated for $\kappa = 0, 0.5$ and 1.}
\label{table_B0bar_BE1}
\end{table}

\begin{table}
\begin{tabular} {|c|c|c|c|c|c|c|c|}
\hline
\multirow{2}{*}{$A$} & $nl$  & \multicolumn{3}{|c|}{$\cal E_B$ (MeV)} & \multicolumn{3}{|c|}{$\Gamma$ (MeV)} \\
\cline{3-8}
 & & $\kappa$ =0 & $\kappa$ = 0.5 & $\kappa$ =1 & $\kappa$ =0 & $\kappa$ =  0.5 & $\kappa$ =1 \\
\hline
\multirow{1}{*}{$^{12}_{6}$C} & 1s  & -0.573  & -0.128  & -- & 0.0 & 2.368 & --  \\
\hline
\multirow{1}{*}{$^{16}_{8}$O} & 1s  & -1.84  & -1.469  & -0.462 & 0.0 & 4.071 & 8.731  \\
\hline
\multirow{1}{*}{$^{40}_{20}$Ca} & 1s  & -11.209  & -10.949  & -10.24 & 0.0 & 10.803 & 21.964  \\
\hline
\multirow{5}{*}{$^{208}_{82}$Pb} & 1s  & -28.644  & -28.509  & -28.123 & 0.0 & 17.663 & 35.389  \\
\cline{2-8}
& 1p  & -21.387  & -21.202  & -20.688 & 0.0 & 16.485 & 33.146  \\
\cline{2-8}
&  1d  & -12.892  & -12.621  & -11.9 & 0.0 & 14.939 & 30.257  \\
\cline{2-8}
&  2s  & -10.123  & -9.746  & -8.799 & 0.0 & 13.919 & 28.501  \\
\cline{2-8}
&  2p  & -0.991  & --  & -- & 0.0 & -- & --  \\
\hline
\end{tabular}
\caption{Values of binding energies, $\cal E_B $, and full decay width, $\Gamma$ for $\bar{K^0}$ mesons in four nuclei with different mass number $A$ are tabulated for $\kappa = 0, 0.5$ and 1.}
\label{table_K0bar_BE1}
\end{table}

\begin{figure}
\centering
\includegraphics[width=15cm, height=8cm]{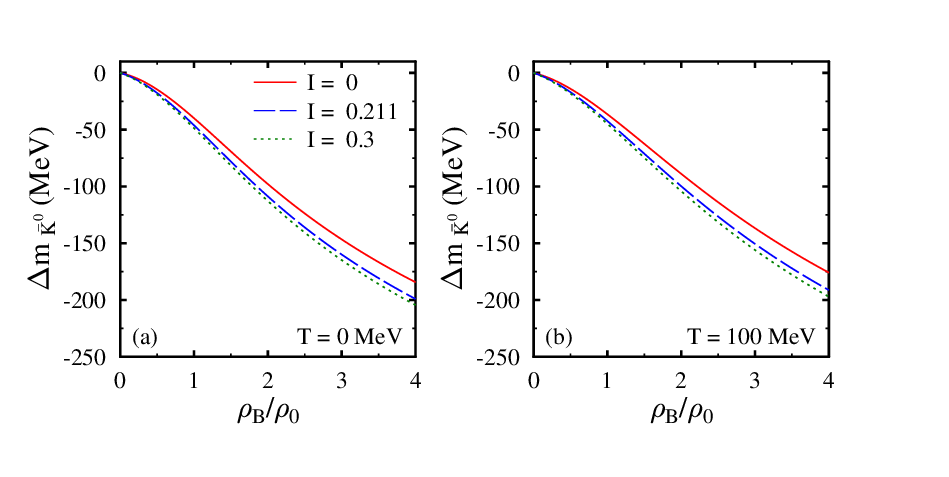}
%\vspace{-2cm}
\caption{ The mass shift of $\bar{K^{0}}$   mesons is shown as a function of density $\rho_B$ (in units of nuclear saturation density $\rho_0$) of nuclear medium for isospin asymmetry, $I = 0, 0.211$ and $0.3$.
Results are shown for temperatures $T = 0 $ (subplot (a)) and $100$ MeV (subplot (b)). \label{fig:K0mass}}
\end{figure}

\begin{figure}
\centering
\includegraphics[width=12cm, height=12cm]{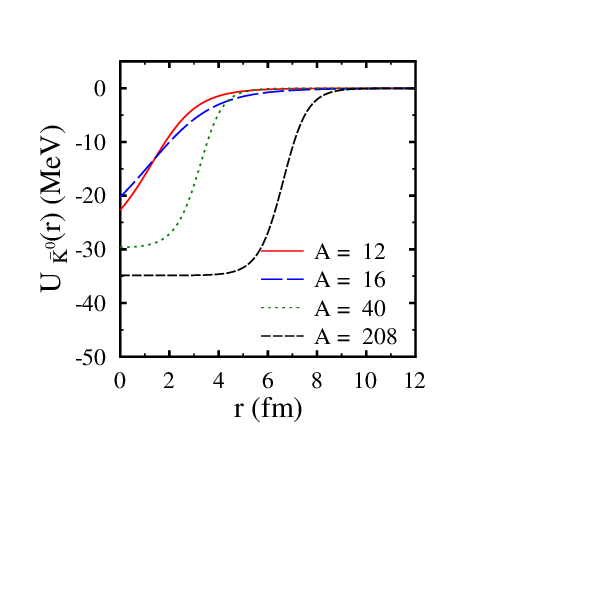}
\vspace{-3cm}
\caption{ The real optical potential $U(r)$  of  $\bar{K^{0}}$ mesons is shown as a function of distance $r$ from center, for nuclei
$^{12}$C, $^{16}$O, $^{40}$Ca and $^{208}$Pb, with mass number, $A = 12, 16, 40$ and $208$, respectively. \label{fig:K0_optical}}
\end{figure}

\begin{figure}
\centering
\includegraphics[width=15cm, height=16cm]{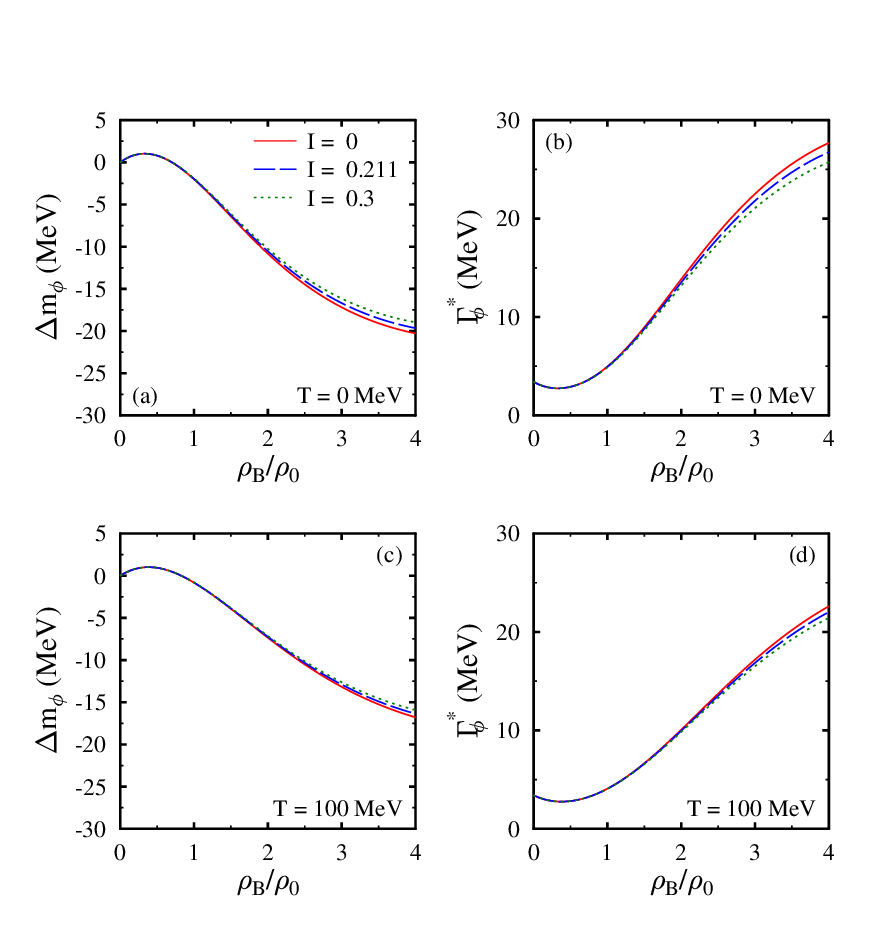}
%\vspace{-2cm}
\caption{The mass shift [in subplots (a) and (c)] and decay width [in subplots (b) and (d)] of $\phi$ mesons are shown as a function of density $\rho_B$ (in units of nuclear saturation density $\rho_0$) of nuclear medium for isospin asymmetry, $I = 0, 0.211$ and $0.3$. Results are shown for temperatures $T = 0 $ and $100$ MeV. \label{fig:phimass}}
\end{figure}

\begin{figure}
\centering
\includegraphics[width=15cm, height=12cm]{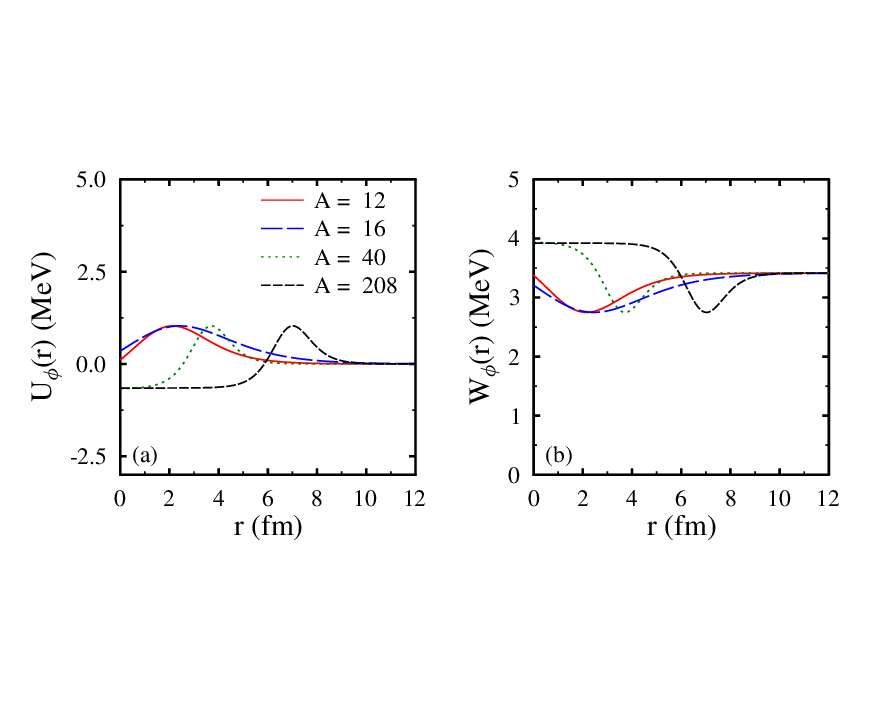}
\vspace{-2cm}
\caption{ The real optical potential $U(r)$  [in subplot (a)] and imaginary optical potential $W(r)$ [in subplot (b)] of $\phi$ mesons are shown as a function of distance $r$ from center, for nuclei
$^{12}$C, $^{16}$O, $^{40}$Ca and $^{208}$Pb,
 with mass number, $A = 12, 16,, 40$ and $208$, respectively. \label{fig:phi_optical}}
\end{figure}

\begin{figure}
\centering
\includegraphics[width=15cm, height=12cm]{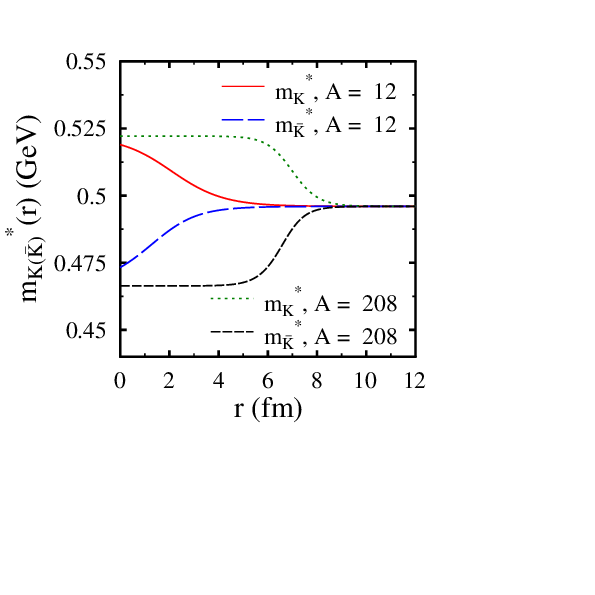}
\vspace{-2cm}
\caption{ The average effective masses
$m_{K}^{*}\left(=\frac{m_{K^{0}}^{*} + m_{K^{+}}^{*}}{2}\right)$ and $m_{\bar{K}}^{*}\left(=\frac{m_{\Bar{K}^{0}}^{*}+m_{K^{-}}^{*}}{2}\right)$
 for kaons and antikaons are plotted as a function of distance $r$ from center, for nuclei
$^{12}$C and $^{208}$Pb,
 with mass number, $A = 12$  and $208$, respectively. \label{fig:Fig_K_optical_r_mkavg}}
\end{figure}
\subsection{Binding energies for   $\bar{K^{0}}$ and $\phi$ mesic nuclei}
Lastly, in this subsection we explore the formation of bound states for $\bar{K^{0}}$ and $\phi$
mesons.
Within the chiral SU(3) model, among the neutral $K^0$ and $\bar{K^0}$ mesons, only latter is observed to undergo mass drop as a function of density of the nuclear medium  \cite{Mishra:2006wy,Mishra:2008kg,Mishra:2008dj} and therefore, we consider $\bar{K^0}$ neutral meson only to explore the  bound state formation. In Fig. \ref{fig:K0mass} we have shown the variation of mass shift for $\bar{K^0}$ mesons as a function of baryon density $\rho_B$, at temperatures $T = 0$ and 100 MeV. The plot of real part of optical potential $U(r)$  as a function of distance $r$ from the center of the given nuclei of mass number $A$ is given for $\bar{K^0}$ meson in Fig. \ref{fig:K0_optical}.
At $I = 0$ and $T = 0$ (100) MeV, the values of mass shift 
for  $\bar{K^0}$ mesons are found to be $-36.90 (-36.43)$ and $-184.31 (-175.87)$ MeV, at $\rho_B = \rho_0$ and $4\rho_0$, respectively.
For a given density of medium, an increase in the isospin asymmetry  enhances the mass drop for $\bar{K^0}$ mesons, which is more appreciable at higher baryon densities.
As can be seen from Table \ref{table_K0bar_BE1}, for the nuclei $^{12}$C,
 $^{18}$O and  $^{40}$Ca,
 the negative values of binding energies are observed only for $1s$ state. Further, in this case when the value of $\kappa$ is considered finite, it may be difficult to observe the
 bound state of $\bar{K^0}$ meson with these nuclei
 due to comparable or larger value of decay width $\Gamma$. 
 
 As we discussed earlier in 
 Sec. \ref{Subsec_phi},
 the mass shift and decay width of $\phi$ meson are obtained in the present work by calculating the self energy, which involve the interactions of $\phi$ mesons with $K$ and $\bar{K}$, at one loop level.
 In Fig. \ref{fig:phimass} 
 we have plotted the mass shift and decay width of $\phi$ mesons as a function of $\rho_B/\rho_0$. 
At temperature $T =0$ MeV, the values of mass shift $\Delta m_{\phi}^*$ and decay width $\ \Gamma_{\phi}^*$ at baryon density $\rho_B = \rho_0 (4\rho_0)$ 
 are observed to be $-2.00 (-20.28)$ and $4.96 (27.68)$ MeV, respectively, for the value of cutoff parameter, $\Lambda_c = 3000$ MeV.
In Fig. \ref{fig:phi_optical} the real and imaginary optical potentials, $U_\phi(r)$ and $W_\phi(r)$ of $\phi$ mesons
are plotted as a function of distance $r$ for the four nuclei ($A = 12, 16, 40$ and $208$).
As we observed very small negative mass shift for the $\phi$ mesons, the bound states are not possible for these mesons within the present calculations where the in-medium masses of $K$ and $\bar{K}$ mesons are evaluated using the chiral SU(3) model.
In the chiral SU(3) model, the properties of kaons and antikaons are modified differently in the nuclear medium. The effective mass of kaons increases whereas the mass of antikaons decrease as the density of the medium is changed from zero to finite value \cite{Mishra:2006wy,Mishra:2008kg,Mishra:2008dj}.  From Fig. \ref{fig:phi_optical}, 
we observe that the real part of the optical potential remains mostly repulsive. This is because of dominance of repulsive contributions of kaons as a function of $r$ over the attractive contribution of antikaons to the in-medium masses of $\phi$ mesons. As discussed in Sec. \ref{Subsec_phi}, average masses $m_{K}^{*}\left(=\frac{m_{K^{0}}^{*} + m_{K^{+}}^{*}}{2}\right)$ and $m_{\bar{K}}^{*}\left(=\frac{m_{\Bar{K}^{0}}^{*}+m_{K^{-}}^{*}}{2}\right)$ are used in the
calculations of self-energy of $\phi$ mesons
given by Eq. (\ref{Eq_real_self2_phiK}). 
As shown in Fig.\ref{fig:Fig_K_optical_r_mkavg}, average mass
$m_{K}^{*}$ remains repulsive (in-medium mass greater than vacuum mass ($\sim$ 494 MeV)) as a function of $r$ whereas $m_{\bar{K}}^{*}$
is attractive (in-medium mass less than vacuum mass ($\sim$ 494 MeV)). For certain range of $r$ values, increased dominance of repulsive contribution causes a bump in real part of optical potential of $\phi$ mesons and a well in the imaginary part.
In Ref. \cite{Cobos-Martinez:2017woo}, the $\phi$ meson bound states are investigated considering the modification of kaons and antikaons within the QMC model. The in-medium masses of kaons and antikaons are considered same in the nuclear medium and are observed to decrease as a function of baryon density $\rho_B$. Thus, the mass drop in the $\phi$ mesons is also observed to be large
as compared to our present calculations. For example, at $\rho_B = \rho_0$, $\Delta m_\phi$ is observed to be 
$\approx - 20$ MeV for the cuttoff $\Lambda_c =  2000$ MeV. This larger mass drop in the $\phi$ meson masses also leads to the possibility of bound states in the calculations of Ref. \cite{Cobos-Martinez:2017woo}.
However, as discussed above, 
in the chiral SU(3) model considered
in the present calculations, the in-medium mass of kaons increases whereas the mass of antikaon decreases as a function of density of nuclear medium. Actually, the effective masses of kaons and antikaons in the chiral SU(3) model are modified in the nuclear medium through the Weinberg Tomozawa term, scalar meson exchange term and range terms
\cite{Mishra:2006wy,Mishra:2008kg,Mishra:2008dj}.
As a function of density of the nuclear medium, the leading order Weinberg Tomozawa term gives repulsive contributions to kaons and attractive for antikaons. Overall this leads to an  increase in the mass of kaons
and a decrease in the mass of antikaons as function of $\rho_B$. Since the in-medium masses of $\phi$ mesons are modified through the medium modifications of kaons and antikaons, their opposite behavior as a function of density leads to small mass shift in the mass of $\phi$ mesons in the present calculations.

%
%\begin{table}
%\begin{tabular} {|c|c|c|c|}
%\hline
%\multirow{1}{*}{A} & nl  & \multicolumn{1}{|c|}{B.E.} & \multicolumn{1}{|c|}{-$\Gamma$/2} \\
%\cline{3-4}
%\hline
%\multirow{3}{*}{$^{12}_{6}{C}$} & 1s  & 0.066 & 0.002  \\
%\cline{2-4}
%& 1p  & 0.065  & 0.0  \\
%\cline{2-4}
%&  1d  & 0.065  & 0.0 \\
%\hline
%\multirow{3}{*}{$^{16}_{8}{O}$} & 1s  & 0.065  & 0.003  \\
%\cline{2-4}
%& 1p  & 0.064  & 0.0  \\
%\cline{2-4}
%&  1d  & 0.064  & 0.0  \\
%\hline
%\multirow{3}{*}{$^{40}_{20}{Ca}$} & 1s  & 0.062  & 0.009 \\
%\cline{2-4}
%& 1p  & 0.062  & 0.0  \\
%\cline{2-4}
%&  1d  & 0.062  & 0.0 \\
%\hline
%\multirow{3}{*}{$^{208}_{82}{Pb}$} & 1s  & 0.074  & 0.023 \\
%\cline{2-4}
%& 1p  & 0.06 & 0.001 \\
%\cline{2-4}
%&  1d  & 0.06  & 0.0  \\
%\hline
%\end{tabular}
%\caption{$\phi$}
%\end{table}

\section{Summary}
\label{sec:summary} 
In summary, we investigated the
possibility of the formation of bound states for neutral mesons 
$\eta, \eta^{'}$, $D^{0}, \bar{D^0}$, $B^0, \bar{B^0}$, $\bar{K^0}$ and $\phi$ mesons with   nuclei,
$^{12}$C, $^{16}$O, $^{40}$Ca
and $^{208}$Pb. The binding energy for different mesons were calculated by solving the Klein Gordon equation
considering the complex potential.
The imaginary potential is introduced to understand the impact of absorption of mesons in the nuclei on the formation of bound states.  The mass shifts for different mesons considered in the present work are used to obtain the real and imaginary optical potentials for nuclei of different mass numbers. 
Except for vector $\phi$ mesons, the imaginary optical potential 
for all other mesons is introduced 
in the current work through phenomenological prescription in terms of a parameter $\kappa$
following some previous studies \cite{Cobos-Martinez:2023hbp}. 

For the study of bound states for pseudoscalar $\eta$ and $\eta^{'}$ mesons within the chiral SU(3) model, we considered the mixing between the states $\eta_{8}$ and $\eta_0$.  We observed appreciable mass drop for the $\eta$ mesons whereas $\eta^{'}$ mass is observed to decrease very less. This leads to the possibility of bound states of $\eta$ mesons with the nuclei considered in this work, whereas for the $\eta^{'}$ meson no bound state is observed. 
Using the chiral SU(3) model, generalized to SU(4) sector, we also elaborated on the formation of bound states for $D^0$ and $\bar{D^0}$ mesons. For both $D^0$ and $\bar{D^0}$ mesons the mass drop observed at nuclear saturation density is such that it lead to formation of bound states.
Among $D^0$ and $\bar{D^0}$, former is observed to undergo larger mass drop and this lead to the formation of bound states with higher states also. For the case of 
$B^0$ and $\bar{B^0}$ mesons also the obtained negative mass shift lead to the formation of bound states for lower and excited states, especially for nuclei with larger mass number $A$. 
In our present work we also observed that the pseudoscalar $\bar{K^0}$ 
mesons can form the bound states 
as negative mass shift is observed
within the chiral SU(3) model.
For the case of vector $\phi$ mesons 
the mass drop obtained in current work is not large enough to form their bound states  with the light and heavy nuclei.
 In Ref. \cite{Mondal:2024vyt}, meson-nucleus bound states are studied using the QMC model. 
 In the QMC model, quarks are treated as fundamental degrees of freedom and are confined inside the 
 nucleon bag. These quarks confined inside the nucleons interact through the exchange of scalar fields $\sigma$ and $\delta$ and the vector fields $\omega$ and $\rho$. 
In the chiral SU(3) hadronic mean field model, hadrons are treated as point particle. As discussed before, the properties of nucleons in the
chiral SU(3) hadronic mean field model are modified through the exchange of scalar mesons, $\sigma, \zeta$ and $\delta$ and  vector mesons $\omega$ and $\rho$.
In the QMC model, the effective masses of different mesons in the nuclear medium are calculated through the energy of static bag in which quark and antiquarks are confined. On the otherhand, as explained in detail in Sec. \ref{sec:potential}, in chiral hadronic model, the interaction Lagrangian density, describing the interactions of mesons under study with the nucleons of the medium are written.
The dispersion relations are obtained from the interaction Lagrangian to calculate the optical potentials of mesons in the nuclear medium. In Ref. \cite{Mondal:2024vyt}, the possibilities of bound states for $K$, $D$ and $B$ mesons were explored whereas in our current work,
bound states of $\eta, \eta^{\prime}$ and $\phi$ mesons with different nuclei were also considered.

We have not considered the effect of finite density and temperature of the nuclear medium on the binding energy of mesic nuclei. An estimate of this can be obtained considering following picture:  consider the formation of mesic nuclei in a nuclear  medium at finite temperature, the medium effects can be incorporated  by replacing the vacuum mass of mesons with the in-medium value in the calculation
of binding energy (appearing in definitions of reduced mass $\mu$ [in Eq. (\ref{Eq_KG_Coord1})] and  the real  and imaginary part of optical potentials [Eqs. (\ref{Eq_real_opt1}) and \ref{Eq_imag_decay_W1})] at given density and temperature.
As an example, consider the formation of $D^{0}$ mesic nuclei in the nuclear medium $\rho_B = \rho_0$ and $T = 100$ MeV, at which
the mass of $D^{0}$ meson is found to be
$1799.57$ MeV. Using this value, the mass shift
$\Delta m_\psi (\rho_0)$ at center of $^{208}_{82}$Pb  nuclei  is found to be $-42.29$ MeV.
This can be compared with the value $-64.98$ MeV which was calculated without considering the impact of nuclear medium on formation of bound states  and mass of $D^{0}$ in free space was 1864 MeV
 (see Fig.\ref{fig:D_optical}(a)).
Using the modifications due to finite density and temperature, the values of binding energy for $1s$ and $1p$ states change to $-40.10$ and $-37.41$ MeV and can be compared values $-49.133$ and $-43.096$ MeV given in Table \ref{table_D0_BE1}. 
Thus, we observe a reduction in the magnitude of binding energy at finite density and temperature of the medium.
The decrease in the magnitude of binding energy is also expected for other cases where bound states 
 with different mesons are formed due to attractive interactions in the medium.
The many-body effects which  may impact the density distributions
of protons and neutrons inside the nuclei and may also cause
additional attractive interaction and absorption
 \cite{Itahashi:2012ut} will be studied in future work.
The present investigations on the formation of meson-nuclei bound states 
will be significant for the experiments of $\bar{\text{P}}$ANDA at FAIR, WASA at COSY and J-PARC facilities.

%\begin{thebibliography}{100}
%\end{thebibliography}

%apsrev4-2.bst 2019-01-14 (MD) hand-edited version of apsrev4-1.bst
%Control: key (0)
%Control: author (8) initials jnrlst
%Control: editor formatted (1) identically to author
%Control: production of article title (0) allowed
%Control: page (0) single
%Control: year (1) truncated
%Control: production of eprint (0) enabled
\begin{thebibliography}{0}%
\makeatletter
\providecommand \@ifxundefined [1]{%
 \@ifx{#1\undefined}
}%
\providecommand \@ifnum [1]{%
 \ifnum #1\expandafter \@firstoftwo
 \else \expandafter \@secondoftwo
 \fi
}%
\providecommand \@ifx [1]{%
 \ifx #1\expandafter \@firstoftwo
 \else \expandafter \@secondoftwo
 \fi
}%
\providecommand \natexlab [1]{#1}%
\providecommand \enquote  [1]{``#1''}%
\providecommand \bibnamefont  [1]{#1}%
\providecommand \bibfnamefont [1]{#1}%
\providecommand \citenamefont [1]{#1}%
\providecommand \href@noop [0]{\@secondoftwo}%
\providecommand \href [0]{\begingroup \@sanitize@url \@href}%
\providecommand \@href[1]{\@@startlink{#1}\@@href}%
\providecommand \@@href[1]{\endgroup#1\@@endlink}%
\providecommand \@sanitize@url [0]{\catcode `\\12\catcode `\$12\catcode
  `\&12\catcode `\#12\catcode `\^12\catcode `\_12\catcode `\%12\relax}%
\providecommand \@@startlink[1]{}%
\providecommand \@@endlink[0]{}%
\providecommand \url  [0]{\begingroup\@sanitize@url \@url }%
\providecommand \@url [1]{\endgroup\@href {#1}{\urlprefix }}%
\providecommand \urlprefix  [0]{URL }%
\providecommand \Eprint [0]{\href }%
\providecommand \doibase [0]{https://doi.org/}%
\providecommand \selectlanguage [0]{\@gobble}%
\providecommand \bibinfo  [0]{\@secondoftwo}%
\providecommand \bibfield  [0]{\@secondoftwo}%
\providecommand \translation [1]{[#1]}%
\providecommand \BibitemOpen [0]{}%
\providecommand \bibitemStop [0]{}%
\providecommand \bibitemNoStop [0]{.\EOS\space}%
\providecommand \EOS [0]{\spacefactor3000\relax}%
\providecommand \BibitemShut  [1]{\csname bibitem#1\endcsname}%
\let\auto@bib@innerbib\@empty
%</preamble>
\end{thebibliography}%


\begin{thebibliography}{100}

%\cite{Hayano:2008vn}
\bibitem{Hayano:2008vn}
R.~S.~Hayano and T.~Hatsuda,
%``Hadron properties in the nuclear medium,''
Rev. Mod. Phys. \textbf{82}, 2949 (2010).
% %doi:10.1103/RevModPhys.82.29%49
%[arXiv:0812.1702 [nucl-ex]].
%248 citations counted in %INSPIRE as of 30 Jul 2024


%\cite{Yasui:2012rw}
\bibitem{Yasui:2012rw}
S.~Yasui and K.~Sudoh,
%``$\bar{D}$ and $B$ mesons in nuclear medium,''
Phys. Rev. C \textbf{87},  015202 (2013).
%doi:10.1103/PhysRevC.87.015202
%[arXiv:1207.3134 [hep-ph]].
%59 citations counted in INSPIRE as of 30 Jul 2024

%\cite{Lee:2023ofg}
\bibitem{Lee:2023ofg}
S.~H.~Lee,
%``Chiral Symmetry Breaking and the Masses of Hadrons: A Review,''
Symmetry \textbf{15},  799 (2023).
%doi:10.3390/sym15040799[arXiv:2303.14415 [hep-ph]].
%5 citations counted in INSPIRE as of 30 Jul 2024

%\cite{Leupold:2009kz}
\bibitem{Leupold:2009kz}
S.~Leupold, V.~Metag and U.~Mosel,
%``Hadrons in strongly interacting matter,''
Int. J. Mod. Phys. E \textbf{19}, 147 (2010).
%doi:10.1142/S0218301310014728[arXiv:0907.2388 [nucl-th]].
%205 citations counted in %INSPIRE as of 30 Jul 2024

%\cite{Metag:2017yuh}
\bibitem{Metag:2017yuh}
V.~Metag, M.~Nanova and E.~Y.~Paryev,
%``Meson\textendash{}nucleus potentials and the search for meson\textendash{}nucleus bound states,''
Prog. Part. Nucl. Phys. \textbf{97}, 199 (2017).

	
	%\cite{Cobos-Martinez:2017woo}
	\bibitem{Cobos-Martinez:2017woo}
	J.~J.~Cobos-Mart\'\i{}nez, K.~Tsushima, G.~Krein and A.~W.~Thomas,
%	``$\Phi$-meson--nucleus bound states,''
	Phys. Rev. C \textbf{96}, 035201 (2017). 


%\cite{Tsushima:1998ru}
\bibitem{Tsushima:1998ru}
K.~Tsushima, D.~H.~Lu, A.~W.~Thomas, K.~Saito and R.~H.~Landau,
%``Charmed mesic nuclei,''
Phys. Rev. C \textbf{59}, 2824 (1999).

\bibitem{Nagahiro:2006dr}
H.~Nagahiro, M.~Takizawa and S.~Hirenzaki,
%``eta- and eta-prime-mesic nuclei and U(A)(1) anomaly at finite density,''
Phys. Rev. C \textbf{74}, 045203 (2006).
%doi:10.1103/PhysRevC.74.04520%3
%[arXiv:nucl-th/0606052 %[nucl-th]].
%128 citations counted in %INSPIRE as of 27 Jul 2024




%\cite{Nagahiro:2011fi}
\bibitem{Nagahiro:2011fi}
H.~Nagahiro, S.~Hirenzaki, E.~Oset and A.~Ramos,
%``eta-prime nucleus optical potential and possible eta-prime bound states,''
Phys. Lett. B \textbf{709}, 87 (2012).
%-92
%doi:10.1016/j.physletb.2012.01.061
%[arXiv:1111.5706 [hep-ph]].
%79 citations counted in INSPIRE as of 14 Jun 2024

\bibitem{Garcia-Recio:2010fiq}
C.~Garcia-Recio, J.~Nieves and L.~Tolos,
%``D mesic nuclei,''
Phys. Lett. B \textbf{690}, 369 (2010).

%\cite{Haider:1986sa}
\bibitem{Haider:1986sa}
Q.~Haider and L.~C.~Liu,
%``Formation of an $\eta$ Mesic Nucleus,''
Phys. Lett. B \textbf{172}, 257 (1986)
%-260
%doi:10.1016/0370-2693(86)90846-4
%251 citations counted in INSPIRE as of 13 Jun 2024


%\cite{Cheng:1987yw}
\bibitem{Cheng:1987yw}
W.~K.~Cheng, T.~T.~S.~Kuo and G.~L.~Li,
%``On Formation of $\eta$ Meson - Nucleus Bound States,''
Phys. Lett. B \textbf{195}, 515 (1987). 
 



%\cite{Nagahiro:2012aq}
\bibitem{Nagahiro:2012aq}
H.~Nagahiro, D.~Jido, H.~Fujioka, K.~Itahashi and S.~Hirenzaki,
%``Formation of $\eta^\prime(958)$-mesic nuclei by ($p,d$) reaction,''
Phys. Rev. C \textbf{87} (2013)  045201
%doi:10.1103/PhysRevC.87.045201
%[arXiv:1211.2506 [nucl-th]].
%89 citations counted in INSPIRE as of 14 Jun 2024


 
\bibitem{Cobos-Martinez:2023hbp}
J.~J.~Cobos-Martinez and K.~Tsushima,
%``\ensuremath{\eta} and \ensuremath{\eta}' mesons in nuclear matter and nuclei,''
Phys. Rev. C \textbf{109}, 2 (2024).
%doi:10.1103/PhysRevC.109.025202
%[arXiv:2308.07836 [nucl-th]].
%2 citations counted in INSPIRE as of 15 Jun 2024

%\cite{Ikeno:2024slo}
\bibitem{Ikeno:2024slo}
N.~Ikeno, \textit{et.al.},
%``Feasibility of the observation of $\eta^{\prime}$ mesic nuclei in the semi-exclusive $^{12}$C($p, dp$) reaction,''
arXiv:2406.06058.
%0 citations counted in INSPIRE as of 15 Jun 2024

%\cite{Kelkar:2013lwa}
\bibitem{Kelkar:2013lwa}
N.~G.~Kelkar, K.~P.~Khemchandani, N.~J.~Upadhyay and B.~K.~Jain,
%``Interaction of eta mesons with nuclei,''
Rept. Prog. Phys. \textbf{76}, 066301 (2013).
%doi:10.1088/0034-4885/76/6/066301
%[arXiv:1306.2909 [nucl-th]].
%67 citations counted in INSPIRE as of 15 Jun 2024


%\cite{Garcia-Recio:2002xog}
\bibitem{Garcia-Recio:2002xog}
C.~Garcia-Recio, J.~Nieves, T.~Inoue and E.~Oset,
%``eta bound states in nuclei,''
Phys. Lett. B \textbf{550}, 47 (2002).
%doi:10.1016/S0370-2693(02)02960-X
%[arXiv:nucl-th/0206024 [nucl-th]].
%113 citations counted in INSPIRE as of 15 Jun 2024


%\cite{Sakai:2022xao}
\bibitem{Sakai:2022xao}
S.~Sakai and D.~Jido,
%``Spectral function of the \ensuremath{\eta}' meson in nuclear medium based on phenomenological models,''
Phys. Rev. C \textbf{107}, 025207 (2023).
%doi:10.1103/PhysRevC.107.025207
%[arXiv:2212.05655 [nucl-th]].
%2 citations counted in INSPIRE as of 15 Jun 2024


%\cite{Hirenzaki:2013txr}
\bibitem{Hirenzaki:2013txr}
S.~Hirenzaki, N.~Ikeno and H.~Nagahiro,
%``Meson Properties at Finite Density from Mesic Atoms and Mesic Nuclei''
% = Recent Topics on $\pi$ and $\eta'(958)$ in Nuclei =,''
PoS \textbf{Hadron2013},  014 (2013)
%doi:10.22323/1.205.0014
%0 citations counted in INSPIRE as of 14 Jun 2024

%\cite{Chrien:1988gn}
\bibitem{Chrien:1988gn}
R.~E.~Chrien \textit{et al.}
%``Search for Bound States of the $\eta$ Meson in Light Nuclei,''
Phys. Rev. Lett. \textbf{60} (1988) 2595.
%-2598
%doi:10.1103/PhysRevLett.60.2595
%122 citations counted in INSPIRE as of 14 Jun 2024

%\cite{Johnson:1993zy}
\bibitem{Johnson:1993zy}
J.~D.~Johnson  \textit{et al.}
%``Search for an eta bound state in pion double charge exchange on O-18,''
Phys. Rev. C \textbf{47},  2571 (1993).
%doi:10.1103/PhysRevC.47.2571
%34 citations counted in INSPIRE as of 14 Jun 2024


%\cite{Sokol:2001hx}
\bibitem{Sokol:2001hx}
G.~A.~Sokol, A.~I.~L'vov and L.~N.~Pavlyuchenko,
%``Discovery of \ensuremath{\eta}-mesic Nuclei,''
%doi:10.1142/9789812704948\_0024,
arXiv:nucl-ex/0111020.

%\cite{Afanasiev:2011zza}
\bibitem{Afanasiev:2011zza}
S.~V.~Afanasiev  \textit{et al.}
%``Search for eta-mesic nuclei in the reaction d + C at JINR,''
Phys. Part. Nucl. Lett. \textbf{8}, 1073 (2011).
%doi:10.1134/S1547477111100025
%17 citations counted in INSPIRE as of 14 Jun 2024


%\cite{WASA-at-COSY:2020bch}
\bibitem{WASA-at-COSY:2020bch}
P.~Adlarson \textit{et al.} [WASA-at-COSY],
%``Search for the $\eta$ mesic $^{3}$He in the $pd \to dp\pi^0$ reaction with the WASA-at-COSY facility,''
Phys. Rev. C \textbf{102},  044322 (2020).
%doi:10.1103/PhysRevC.102.044322
%[arXiv:2007.15494 [nucl-ex]].
%9 citations counted in INSPIRE as of 15 Jun 2024

%\cite{Adlarson:2019haw}
\bibitem{Adlarson:2019haw}
P.~Adlarson,  \textit{et al.}
%``Search for $\eta$ mesic $^3$He with the WASA-at-COSY facility in the $pd \to$ $^3$He 2$\gamma$ and $pd \to$ $^3$He6$\gamma$ reactions,''
Phys. Lett. B \textbf{802}, 135205 (2020).
%doi:10.1016/j.physletb.2020.135205
%[arXiv:1909.10780 [nucl-ex]].
%11 citations counted in INSPIRE as of 15 Jun 2024


%\cite{Wu:2023cfh}
\bibitem{Wu:2023cfh}
Q.~Wu, G.~Xie and X.~Chen,
%``Probing the existence of \ensuremath{\eta} $^{3}$He mesic nucleus with a few-body approach,''
Phys. Lett. B \textbf{850}, 138502 (2024).
%doi:10.1016/j.physletb.2024.138502
%[arXiv:2309.14711 [nucl-th]].
%0 citations counted in INSPIRE as of 15 Jun 2024

%\cite{Bass:2021rch}
\bibitem{Bass:2021rch}
S.~D.~Bass, V.~Metag and P.~Moskal,
%``The \ensuremath{\eta}- and \ensuremath{\eta}'-Nucleus Interactions and the Search for \ensuremath{\eta}, \ensuremath{\eta}'- Mesic States,''
%doi:10.1007/978-981-15-8818-1\_39-1
arXiv:2111.01388 [hep-ph].
%4 citations counted in %INSPIRE as of 27 Jul 2024


%\cite{Khreptak:2023lbh}
\bibitem{Khreptak:2023lbh}
A.~Khreptak, M.~Skurzok and P.~Moskal,
%``Search for \ensuremath{\eta}-mesic nuclei: a review of experimental and theoretical advances,''
Front. in Phys. \textbf{11}, 1186457 (2023).
%doi:10.3389/fphy.2023.1186457
%2 citations counted in INSPIRE as of 13 Jun 2024
 
 


%\cite{Tolos:2013gta}
\bibitem{Tolos:2013gta}
L.~Tolos,
%``Charming mesons with baryons and nuclei,''
Int. J. Mod. Phys. E \textbf{22}, 1330027 (2013).

\bibitem{Tsushima:2011kh}
K.~Tsushima, D.~H.~Lu, G.~Krein and A.~W.~Thomas,
%``$J/\Psi$-nuclear bound states,''
Phys. Rev. C \textbf{83}, 065208 (2011).

%\cite{Hayashigaki:2000es}
\bibitem{Hayashigaki:2000es}
A.~Hayashigaki,
%``Mass modification of D meson at finite density in QCD sum rule,''
Phys. Lett. B \textbf{487}, 96 (2000).

%\cite{Mizutani:2006vq}
%\cite{Mizutani:2006vq}
\bibitem{Mizutani:2006vq}
T.~Mizutani and A.~Ramos,
%``D mesons in nuclear matter: A DN coupled-channel equations approach,''
Phys. Rev. C \textbf{74}, 065201 (2006).
%doi:10.1103/PhysRevC.74.065201[arXiv:hep-ph/0607257[hep-ph]].
%174 citations counted in INSPIRE as of 30 Jul 2024


%\cite{Tolos:2007vh}
\bibitem{Tolos:2007vh}
L.~Tolos, A.~Ramos and T.~Mizutani,
%``Open charm in nuclear matter at finite temperature,''
Phys. Rev. C \textbf{77}, 015207 (2008).

%\cite{Mondal:2023iwe}
\bibitem{Mondal:2023iwe}
A.~Mondal and A.~Mishra,
%``Open strange and open heavy flavor mesons in asymmetric nuclear matter within quark meson coupling model,''
Phys. Rev. C \textbf{109}, 025201 (2024).


%\cite{Kumar:2020kng}
\bibitem{Kumar:2020kng}
R.~Kumar, R.~Chhabra and A.~Kumar,
%``Heavy vector and axial-vector $D$ mesons in hot magnetized asymmetric nuclear matter,''
Eur. Phys. J. A \textbf{56}, 278 (2020).

%\cite{Gubler:2020hft}
\bibitem{Gubler:2020hft}
P.~Gubler, T.~Song and S.~H.~Lee,
%``D meson mass and heavy quark potential at finite temperature,''
Phys. Rev. D \textbf{101}, 114029 (2020).

 %\cite{Suzuki:2015est}
 \bibitem{Suzuki:2015est}
 K.~Suzuki, P.~Gubler and M.~Oka,
% ``D meson mass increase by restoration of chiral symmetry in nuclear matter,''
 Phys. Rev. C \textbf{93},  045209 (2016).
 
 %\cite{Mishra:2008cd}
 \bibitem{Mishra:2008cd}
 A.~Mishra and A.~Mazumdar,
 %``D-mesons in asymmetric nuclear matter,''
 Phys. Rev. C \textbf{79}, 024908 (2009).
 
 %\cite{Kumar:2011ff}
 \bibitem{Kumar:2011ff}
 A.~Kumar and A.~Mishra,
 %``D-mesons and charmonium states in hot isospin asymmetric strange hadronic matter,''
 Eur. Phys. J. A \textbf{47}, 164 (2011).
 
 
 
 





%\cite{Schadmand:2010bi}
\bibitem{Schadmand:2010bi}
S.~Schadmand,
%``Mesons in Nuclei: $\eta$ and $\omega$ mesons,''
Prog. Theor. Phys. Suppl. \textbf{186}, 373 (2010).
 

%\cite{Cobos-Martinez:2020ynh}
\bibitem{Cobos-Martinez:2020ynh}
J.~J.~Cobos-Mart\'\i{}nez, K.~Tsushima, G.~Krein and A.~W.~Thomas,
%``$\eta_{c}$-nucleus bound states,''
Phys. Lett. B \textbf{811}, 135882 (2020).

%\cite{Cobos-Martinez:2022fmt}
\bibitem{Cobos-Martinez:2022fmt}
J.~J.~Cobos-Mart\'\i{}nez, G.~N.~Zeminiani and K.~Tsushima,
%``\ensuremath{\Upsilon} and \ensuremath{\eta}b nuclear bound states,''
Phys. Rev. C \textbf{105},   025204 (2022).

%\cite{Pathak:2014nfa}
\bibitem{Pathak:2014nfa}
D.~Pathak and A.~Mishra,
%``Open bottom mesons in a hot asymmetric hadronic medium,''
Phys. Rev. C \textbf{91}, 045206 (2015).
%doi:10.1103/PhysRevC.91.045206
%[arXiv:1409.0728 [nucl-th]].
%37 citations counted in INSPIRE %as of 30 Jul 2024

%\cite{Mondal:2024vyt}
\bibitem{Mondal:2024vyt}
A.~Mondal and A.~Mishra,
%``Meson-nucleus bound states in quark meson coupling model,''
arXiv:2407.19896 [nucl-th].
%0 citations counted in %INSPIRE as of 30 Jul 2024

\bibitem{KEK-PS-E325:2005wbm}
R.~Muto \textit{et al.} [KEK-PS-E325],
%``Evidence for in-medium modification of the phi meson at normal nuclear density,''
Phys. Rev. Lett. \textbf{98}, 042501 (2007).

%\cite{Ishikawa:2004id}
\bibitem{Ishikawa:2004id}
T.~Ishikawa, \textit{et al.}
%``phi photo-production from Li, C, Al, and Cu nuclei at E(gamma) = 1.5-GeV to 2.4-GeV,''
Phys. Lett. B \textbf{608}, 215 (2005)
 

%\cite{CLAS:2009kjz}
\bibitem{CLAS:2009kjz}
X.~Qian \textit{et al.} [CLAS],
%``The Extraction of phi-N total cross section from d(gamma,pK+ K-)n,''
Phys. Lett. B \textbf{680}, 417 (2009).

%\cite{CLAS:2010pxs}
\bibitem{CLAS:2010pxs}
M.~H.~Wood \textit{et al.} [CLAS],
%``Absorption of the $\omega$ and $\phi$ Mesons in Nuclei,''
Phys. Rev. Lett. \textbf{105}, 112301 (2010).
 
\bibitem{JparcE16} 
\url{http://rarfaxp.riken.go.jp/~yokkaich/paper/jparc-proposal-0604.pdf}

%\cite{Aoki:2015qla}
\bibitem{Aoki:2015qla}
K.~Aoki [J-PARC E16],
%``Study of in-medium mass modification at J-PARC,''
arXiv:1502.00703.

\bibitem{Hatsuda:1991ez}
	T.~Hatsuda and S.~H.~Lee,
%``QCD sum rules for vector mesons in the nuclear medium,''
	Phys. Rev. C \textbf{46}, R34 (1992).
	\bibitem{Hatsuda:1996xt}
	T.~Hatsuda, H.~Shiomi and H.~Kuwabara,
%	``Light vector mesons in nuclear matter,''
	Prog. Theor. Phys. \textbf{95}, 1009 (1996).
\bibitem{Oset:2000eg}
	E.~Oset and A.~Ramos,
%	``Phi decay in nuclei,''
	Nucl. Phys. A \textbf{679}, 616 (2001).
	\bibitem{Klingl:1997tm}
	F.~Klingl, T.~Waas and W.~Weise,
%	``Modification of the phi meson spectrum in nuclear matter,''
	Phys. Lett. B \textbf{431}, 254 (1998).
	\bibitem{Gubler:2015yna}
	P.~Gubler and W.~Weise,
%	``Moments of \ensuremath{\phi} meson spectral functions in vacuum and nuclear matter,''
	Phys. Lett. B \textbf{751}, 396 (2015).

	
%\cite{}
\bibitem{Paryev:2022zkt}
E.~Y.~Paryev,
%``Testing the \ensuremath{\phi}-nuclear potential in pion-induced \ensuremath{\phi} meson production on nuclei near threshold,''
Nucl. Phys. A \textbf{1032}, 122624 (2023).

%\cite{}
\bibitem{Chizzali:2022pjd}
E.~Chizzali, Y.~Kamiya, R.~Del Grande, T.~Doi, L.~Fabbietti, T.~Hatsuda and Y.~Lyu,
%``Indication of a p\textendash{}\ensuremath{\phi} bound state from a correlation function analysis,''
Phys. Lett. B \textbf{848}, 138358 (2024).

%\cite{Kim:2022eku}
\bibitem{Kim:2022eku}
J.~Kim, P.~Gubler and S.~H.~Lee,
%``\ensuremath{\phi} meson properties in nuclear matter from QCD sum rules with chirally separated four-quark condensates,''
Phys. Rev. D \textbf{105},   114053 (2022).

%\cite{Kumar:2020vys}
\bibitem{Kumar:2020vys}
R.~Kumar and A.~Kumar,
%``$\phi$ meson mass and decay width in strange hadronic matter,''
Phys. Rev. C \textbf{102},  045206 (2020).
 

\bibitem{Japrc_nuc1}
\url{https://j-parc.jp/researcher/Hadron/en/pac_0907/pdf/Ohnishi.pdf}

\bibitem{Japrc_nuc2}
\url{https://j-parc.jp/researcher/Hadron/en/pac_1007/pdf/KEK_J-PARC-PAC2010-02.pdf}
 


 %\cite{Yamagata-Sekihara:2015ebw}
 \bibitem{Yamagata-Sekihara:2015ebw}
 J.~Yamagata-Sekihara, C.~Garcia-Recio, J.~Nieves, L.~L.~Salcedo and L.~Tolos,
 %``Formation spectra of charmed meson\textendash{}nucleus systems using an antiproton beam,''
 Phys. Lett. B \textbf{754}, 26 (2016)
%doi:10.1016/j.physletb.2016.01.003 [arXiv:1512.03240 [nucl-th]].
 %12 citations counted in INSPIRE as of 01 Aug 2024   

\bibitem{Papazoglou:1998vr}
	P.~Papazoglou et al.,
%	``Nuclei in a chiral SU(3) %model,''
	Phys. Rev. C \textbf{59}  (1999) 411.
	
	%\cite{Mishra:2008dj}
	\bibitem{Mishra:2008dj}
	A.~Mishra, A.~Kumar, S.~Sanyal and S.~Schramm,
%	``Kaon and antikaon optical potentials in isospin asymmetric hyperonic matter,''
	Eur. Phys. J. A \textbf{41}, 205  (2009).

 

%\cite{Mishra:2006wy}
\bibitem{Mishra:2006wy}
A.~Mishra and S.~Schramm,
%``Isospin dependent kaon and antikaon optical potentials in dense hadronic matter,''
Phys. Rev. C \textbf{74}, 064904 (2006).
 

%\cite{Mishra:2008kg}
\bibitem{Mishra:2008kg}
A.~Mishra, S.~Schramm and W.~Greiner,
%``Kaons and antikaons in asymmetric nuclear matter,''
Phys. Rev. C \textbf{78}, 024901 (2008).

%\cite{Mishra:2014rha}
 \bibitem{Mishra:2014rha}
 A.~Mishra,
% ``Light vector meson masses in strange hadronic matter: A QCD sum rule approach,''
 Phys. Rev. C \textbf{91}, 035201 (2015).%\cite{Mishra:2014rha}
 
  
  
 %\cite{Kumar:2020gpu}
 \bibitem{Kumar:2020gpu}
 R.~Kumar and A.~Kumar,
% ``$\eta$ mesons in hot and dense asymmetric nuclear matter,''
 Phys. Rev. C \textbf{102},   065207 (2020).
 
 %\cite{Tiwari:2022gre}
 \bibitem{Tiwari:2022gre}
 S.~Tiwari, R.~Kumar, M.~Kumari and A.~Kumar,
% ``$\eta $ meson in strange magnetized matter,''
 Eur. Phys. J. Plus \textbf{139},  310 (2024).
  
  %\cite{Kumar:2010gb}
  \bibitem{Kumar:2010gb}
  A.~Kumar and A.~Mishra,
 % ``D mesons and charmonium states in asymmetric nuclear matter at finite temperatures,''
  Phys. Rev. C \textbf{81}, 065204 (2010).
  

  
 
  \bibitem{Dhale:2018plh}
  N.~Dhale, S.~P.~Reddy, A.~C.~Jahan and A.~Mishra,
 % ``Open bottom mesons in asymmetric nuclear matter in presence of strong magnetic fields,''
  Phys. Rev. C \textbf{98}, 015202 (2018).
   
  %\cite{Kumar:2010hs}
  \bibitem{Kumar:2010hs}
  A.~Kumar and A.~Mishra,
 % ``$J/\psi$ and $\eta_{c}$ masses in isospin asymmetric hot nuclear matter: A QCD sum rule approach,''
  Phys. Rev. C \textbf{82}, 045207 (2010).
    
  \bibitem{Mishra:2014gea}
  A.~Mishra and D.~Pathak,
 % ``Bottomonium states in hot asymmetric strange hadronic matter,''
  Phys. Rev. C \textbf{90}, 025201 (2014).
 
\bibitem{Zhang:2016bem}
X.~Zhang and M.~Prakash,
%``Hot and dense matter beyond relativistic mean field theory,''
Phys. Rev. C \textbf{93}, 055805 (2016).
 \bibitem{Constantinou:2016hvf}
 C.~Constantinou, S.~Lalit and M.~Prakash,
 %``Thermal Effects in Dense Matter Beyond Mean Field Theory,''
 Int. J. Mod. Phys. E \textbf{26}, 1740005 (2017).
   
 	\bibitem{Weinberg:1968de}
 	S.~Weinberg,
 %	 ``Nonlinear realizations of chiral symmetry,''
 	Phys. Rev. \textbf{166}, 1568 (1968).
 	\bibitem{Coleman:1969sm}
 	S.~R.~Coleman, J.~Wess and B.~Zumino,
 %	``Structure of phenomenological Lagrangians. 1.,''
 	Phys. Rev. \textbf{177}, 2239 (1969).
 	\bibitem{Bardeen:1969ra}
 	W.~A.~Bardeen and B.~W.~Lee,
 %	``Some considerations on nonlinear realizations of chiral su(3) x su(3),''
 	Phys. Rev. \textbf{177}, 2389 (1969).
 	
%\cite{Ramos:1999ku}
\bibitem{Ramos:1999ku}
A.~Ramos and E.~Oset,
%``The Properties of anti-K in the %nuclear medium,''
Nucl. Phys. A \textbf{671}, 481 (2000).
%doi:10.1016/S0375-9474(99)00846-5
%[arXiv:nucl-th/9906016 [nucl-th]].
%296 citations counted in INSPIRE as of 16 Feb 2025 	
 	
%\cite{Nieves:1993ev}
\bibitem{Nieves:1993ev}
J.~Nieves, E.~Oset and C.~Garcia-Recio,
%``A Theoretical approach to pionic atoms and the problem of anomalies,''
Nucl. Phys. A \textbf{554}, 509 (1993).

%\cite{Kwan:1978zh}
\bibitem{Kwan:1978zh}
Y.~R.~Kwan and F.~Tabakin,
%``Hadronic Atoms in Momentum Space,''
Phys. Rev. C \textbf{18}, 932 (1978).

%\cite{Landau:1982iu}
\bibitem{Landau:1982iu}
R.~H.~Landau,
%``Coupled Bound and Continuum Eigenstates in Momentum Space,''
Phys. Rev. C \textbf{27}, 2191 (1983).


%\cite{Zhong:2006jj}
\bibitem{Zhong:2006jj}
X.~H.~Zhong, G.~X.~Peng, L.~Li and P.~Z.~Ning,
%``eta mesons in nuclear matter,''
Phys. Rev. C \textbf{73}, 015205 (2006).

 %\cite{Tolos:2005ft}
 \bibitem{Tolos:2005ft}
 L.~Tolos, J.~Schaffner-Bielich and H.~Stoecker,
 %``D-mesons: In-medium effects at FAIR,''
 Phys. Lett. B \textbf{635}, 85 (2006).
 %doi:10.1016/j.physletb.2006.02.045
% [arXiv:nucl-th/0509054 [nucl-th]].
 %80 citations counted in INSPIRE as of 02 Feb 2025

\bibitem{Ko:1992tp}
	C.~M.~Ko, P.~Levai, X.~J.~Qiu and C.~T.~Li,
%	``Phi meson in dense matter,''
	Phys. Rev. C \textbf{45}, 1400 (1992).
	
\bibitem{Cobos-Martinez:2017vtr}
	J.~J.~Cobos-Mart\'\i{}nez, K.~Tsushima, G.~Krein and A.~W.~Thomas,
%	``$\phi$ meson mass and decay width in nuclear matter and nuclei,''
	Phys. Lett. B \textbf{771}, 113 (2017).

\bibitem{Krein:2010vp}
	G.~Krein, A.~W.~Thomas and K.~Tsushima,
%	``$J/\Psi$ mass shift in nuclear matter,''
	Phys. Lett. B \textbf{697}, 136 (2011).
	
	\bibitem{Li:1994cj}
		G.~Q.~Li and C.~M.~Ko,
%		``Can dileptons reveal the in-medium properties of vector mesons?,''
		Nucl. Phys. A \textbf{582}, 731 (1995).
		 




%\cite{Jido:2018aew}
\bibitem{Jido:2018aew}
D.~Jido, H.~Masutani and S.~Hirenzaki,
%``Structure of $\eta^{\prime}$ mesonic nuclei in a relativistic mean field theory,''
PTEP \textbf{2019},  053D02 (2019).
%doi:10.1093/ptep/ptz031[arXiv%:1808.10140 [nucl-th]].
%6 citations counted in %INSPIRE as of 31 Jul 2024

%\cite{Barnes:1992ca}
\bibitem{Barnes:1992ca}
T.~Barnes and E.~S.~Swanson,
%``Kaon - nucleon scattering %amplitudes and Z* enhancements from %quark Born diagrams,''
Phys. Rev. C \textbf{49}, 1166 (1994).
 

%\bibitem{Zakout:1999qu}
 %	I.~Zakout et al.,
 %	``Hot hypernuclear matter in the modified quark meson coupling model,''
 %	Phys. Rev. C \textbf{61}, 055208 (2000).
 
 
 %\cite{Itahashi:2012ut}
 \bibitem{Itahashi:2012ut}
 K.~Itahashi, H.~Fujioka, H.~Geissel, R.~S.~Hayano, S.~Hirenzaki, S.~Itoh, D.~Jido, V.~Metag, H.~Nagahiro and M.~Nanova, \textit{et al.}
 %``Feasibility Study of Observing eta' Mesic %Nuclei with (p,d) Reaction,''
 Prog. Theor. Phys. \textbf{128}, 601 (2012).
% doi:10.1143/PTP.128.601
 %[arXiv:1203.6720 [nucl-ex]].
 %81 citations counted in INSPIRE as of 01 Feb %2025
 	
\end{thebibliography}
\end{document}